\PassOptionsToPackage{table}{xcolor}
\documentclass[sigconf,pbalance,nonacm]{acmart} 

\usepackage{xcolor}

\usepackage{graphicx}

\usepackage{placeins} 

\usepackage{xspace}
\usepackage{listings}

\usepackage{algorithm}
\usepackage[noend]{algpseudocode}

\usepackage{booktabs}
\usepackage{tikz}

\usepackage{siunitx}

\usepackage{subcaption}
\usepackage{url}

\definecolor{colnet}{RGB}{230,159,0}
\definecolor{colmain}{RGB}{86,180,233}
\definecolor{colwrap}{RGB}{204,121,167}
\usepackage{microtype}
\usepackage{enumitem}

\usepackage{pgfplots}
\pgfplotsset{compat=newest}

\begin{document}

\title{pSTL-Bench\xspace: A Micro-Benchmark Suite for Assessing Scalability of C++ Parallel STL Implementations}

\author{Ruben Laso}
\email{ruben.laso@tuwien.ac.at}
\affiliation{%
  \institution{Faculty of Informatics\\TU Wien}
  \city{Vienna}
  \country{Austria}
  \postcode{1040}
}
\author{Diego Krupitza}
\email{krupitza@par.tuwien.ac.at}
\affiliation{%
  \institution{Faculty of Informatics\\TU Wien}
  \city{Vienna}
  \country{Austria}
  \postcode{1040}
}
\author{Sascha Hunold}
\email{sascha.hunold@tuwien.ac.at}
\affiliation{%
  \institution{Faculty of Informatics\\TU Wien}
  \city{Vienna}
  \country{Austria}
  \postcode{1040}
}

\begin{abstract}
  Since the advent of parallel algorithms in the C++17 Standard
  Template Library (STL), the STL has become a viable framework for
  creating performance-portable applications. Given multiple existing
  implementations of the parallel algorithms, a systematic,
  quantitative performance comparison is essential for choosing the
  appropriate implementation for a particular hardware configuration.

  In this work, we introduce a specialized set of micro-benchmarks to
  assess the scalability of the parallel algorithms in the STL. By
  selecting different backends, our micro-benchmarks can be used on
  multi-core systems and GPUs.

  Using the suite, in a case study on AMD and Intel CPUs and NVIDIA GPUs,
  we were able to identify substantial performance disparities among
  different implementations, including GCC+TBB, GCC+HPX, Intel's compiler
  with TBB, or NVIDIA's compiler with OpenMP and CUDA.
\end{abstract}

\keywords{Performance Portability, C++, Standard Template Library, Threading Building Blocks, OpenMP, CUDA}

\maketitle

\section{Introduction}
\label{sec:intro}

Writing efficient, parallel applications is notoriously hard, but writing 
performance-portable, efficient, parallel applications is harder.
In the last decades, several types of parallel architectures (like GPUs or Xeon Phis) were introduced, which often
required a complete rewrite of the parallelization approach to make applications efficient. To overcome the
problem of having to combine several paradigms such as CUDA~\cite{Nickolls2008cuda}, OpenMP~\cite{dagum1998openmp},
or MPI~\cite{MPI1994} to write efficient programs, several frameworks, mainly using C$++$\xspace, such as
Kokkos~\cite{Trott2022} or Raja~\cite{Beckingsale2019}, were proposed to
allow scientists to write performance-portable applications.
These frameworks allow for an efficient execution of parallel applications using different hardware
architectures, i.e.\xspace, the same program can run on one or more GPUs as well as on multi-core CPUs.

With the advent of C$++$17\xspace, parallel versions of the C$++$\xspace Standard Template Library (STL) were standardized, which allows ISO C$++$\xspace parallel programs to be performance portable~\cite{LinDM22}.
Several works have already compared the resulting performance of various performance-portability layers~\cite{Hammond2019,Artigues2020}.
However, their focus lay on comparing full applications or mini-apps, where significant parts of a rewritten program may significantly influence the resulting performance.

In this work, we set out to devise a set of micro-benchmarks to assess the 
performance of the individual parallel STL algorithms found in C$++$\xspace in a quantitative manner.
Since different compiler frameworks provide competing implementations
of the STL, our goal is to capture the current state-of-the-art of the performance 
of parallel STL implementations. We compare several combinations
of compiler frameworks, including GCC, Intel OneAPI compiler, and NVIDIA\xspace HPC SDK, and backends 
like Intel's Threading Building Blocks (TBB), High-Performance ParalleX (HPX), and OpenMP\xspace.

In particular, we make the following contributions:
\begin{enumerate}
    \item We introduce the benchmark suite pSTL-Bench\xspace, which is an extensible set of micro-benchmarks to assess the performance of parallel STL algorithms on different parallel architectures (multi-cores, GPUs).
    \item Using the suite, we conduct a study over a selection of algorithms comparing the performance achieved on current multi-core architectures by different compiler frameworks and backends implementing the parallel STL.
\end{enumerate}

The remainder of the paper is structured as follows.
In Section\xspace~\ref{sec:rel_work}, we give an overview of the field by summarizing the related work
and current state of the art.  Section\xspace~\ref{sec:microbench} introduces 
the specifics of our proposed set of micro-benchmarks.
In Section\xspace~\ref{sec:exp_eval} we detail how the experiments were carried
out before we show and analyze the experimental results in 
Section\xspace~\ref{sec:experimental-results}.
Finally, we draw conclusions from the
findings in~Section\xspace~\ref{sec:conclusions} and outline future work.

\section{Related Work}
\label{sec:rel_work}

Allowing for performance portability has always been a goal for programmers.
This is especially true for developers on HPC systems, as novel HPC
systems often provide new hardware architectures for which no efficient software solutions
exist yet (cf.\xspace Jack Dongarra's interview when receiving the ACM A.M. Turing Award~\cite{Hoffmann2022}). The Message Passing Interface (MPI) is one of the standards that
enables scientists to write efficient, parallel programs that are also portable across architectures. However, MPI has its limits, especially when it comes to efficiently programming multi-core systems or accelerators, such as GPUs or Xeon Phis.

For the reasons mentioned above, several parallel programming frameworks striving for performance portability were introduced. Note that the term ``performance portability'' may have completely different notions or definitions~\cite{Pennycook2019,Pennycook2021a}.
Contrary to Pennycook et~al.\xspace's restrictive definition, in this work, we consider a program to be performance-portable if it can be executed efficiently on different architectures.
We call a program efficient if it performs within a threshold of a specialized program, i.e.\xspace,
a scan algorithm implemented using a performance portability framework performs equally well on an NVIDIA\xspace GPU as the best-known CUDA implementation.

Our work focuses on micro-benchmarking algorithms from the ISO C$++$\xspace standard library.
There are several other feature-rich C$++$\xspace frameworks that offer performance portability.
For instance, Kokkos~\cite{Trott2022} and Raja~\cite{Beckingsale2019} are one of the most prominent
libraries that can be used to program portable HPC applications using different backends, such as SYCL~\cite{Reyes2020}, HPX~\cite{Kaiser2020}, or OpenMP. 
Although Kokkos may use SYCL as a backend, SYCL itself is a C++ programming model that offers portability between heterogeneous compute resources. Similar to Kokkos, SYCL is a C++ abstraction layer that supports a range of processor architectures and accelerators~\cite{Hammond2019}.
For GPUs, the Thrust library~\cite{BELL2012359} provides a high-level interface for STL to C++ programmers.
Thrust is used as a backend in the C++ STL implementation of NVIDIA\xspace{}'s High-Performance Computing (HPC) Software Development Kit.

HPX~\cite{Kaiser2020} is another C++ abstraction layer for
developing efficient parallel and concurrent applications. 
HPX can be used to write distributed applications using an
Active Global Address Space (AGAS), which is an extension to PGAS for
transparently moving global objects in between compute
nodes~\cite{Kaiser2020}.  SYCL, HPX, and Kokkos can also be used
together to combine their strengths~\cite{Daiss2023}.

The benchmark suite SYCL-Bench~\cite{Lal2020} is a collection of
different programs to analyze the performance of various SYCL
implementations. Besides micro-benchmarks, SYCL-Bench also features
real-world application benchmarks, such as 2DConvolution, as well as
so-called runtime benchmarks, such as a DAG benchmark with many
independent tasks.  Pennycook~et~al.\xspace presented an analysis of the
terminology used in SYCL and ISO C++ to shed light on the relationship
between C++ concepts in both programming models~\cite{Pennycook2023}.

This work was inspired by Lin~et~al.\xspace~\cite{LinDM22}, who
examined the ISO C++ capabilities for creating performance portable
applications. In their work, they developed an ISO C++ parallel
version of the BabelStream benchmark~\cite{Deakin2018} and also
adapted real-world applications, such as MiniBude, to utilize the
parallel STL of ISO C++.

In the context of the present work, we will evaluate different
implementations of the parallel STL for multi-core systems. That
collection includes the HPX~\cite{Kaiser2020} library discussed
above. We also analyze the performance of the GNU's parallel STL, 
based on MCSTL~\cite{Singler2007} which uses
OpenMP\xspace for parallelization and can be seen as the original parallel
implementation of the STL. Currently, the GNU compiler and the Intel
compiler use the Intel Threading Building Blocks
(TBB)~\cite{Reinders2007} for implementing the parallel C++ STL,
whereas NVIDIA\xspace relies on several technologies, primarily OpenMP\xspace~\cite{dagum1998openmp} and CUDA~\cite{Nickolls2008cuda}, for its parallel implementation of the STL.

Compared to the previous benchmarks, pSTL-Bench\xspace targets the
scalability of the individual building blocks, in the same spirit as the
OMPTB~\cite{Hunold2022} or the EPCC OpenMP microbenchmark suite~\cite{Bull2012} 
for benchmarking OpenMP\xspace primitives.

\section{Micro-Benchmark Suite pSTL-Bench\xspace}
\label{sec:microbench}

We now present the micro-benchmark suite pSTL-Bench\xspace.

\subsection{Idea and Goals}

As mentioned previously, our primary goal for developing pSTL-Bench\xspace is to assess the scalability and the efficiency of the different implementations of the parallel C$++$\xspace STL.
Specifically, we aim to achieve several objectives.
First, we would like to determine the sweet-spot of problem size for each parallel STL algorithm, i.e.\xspace, how large a problem has to be such that utilizing the parallel version is advantageous.
Second, we aim to ascertain the maximum number of cores that can be effectively utilized by various parallel algorithms, many of which are memory-bound.
Finally, we would like to examine the run-time\xspace achievable with each of the parallel STL implementations and backends.
We intend to assess the run-time\xspace performance achievable with each of the parallel STL implementations and backends. 
To facilitate this comparison of performance across different standard library implementations, we have compiled these benchmarks into a suite named pSTL-Bench\xspace.\footnote{Code can be provided on request.}

\subsection{Description of Benchmark Kernels}

For the sake of conciseness, we select and evaluate five algorithms of different computational structure
in this paper. The \verb|find| algorithm performs a linear search on the input array, while the \verb|for_each| call performs a  map operation on each element of the input array in parallel. The 
\verb|reduce| call represents a parallel reduction operation, needed for map-reduce type programming.
The \verb|inclusive_scan| represents a typical, parallel prefix-sum operation, 
and \verb|sort| was selected to include a parallel sorting function into the set of test algorithms.

A precise description of each of the algorithms is shown below:
\begin{description}
    \item \verb|find|: Given an array of $n$ elements, $v = [1, 2, \dots, n]$, find a random element $v_i$, such that $v_i \in v$.

    \item \verb|for_each|: Given an array of $n$ elements, $v = [1, 2, \dots, n]$, compute for each element $\min\{\sin(v_i), \tan(v_i)\}$.

    \item \verb|inclusive_scan|: Given an array of $n$ elements, $v = [1, 2, \dots, n]$, compute the result of inclusive sum, $r$, such that $r_i = \sum_{j=1}^{j} v_j$.

    \item \verb|reduce|: Given an array of $n$ elements, $v = [1, 2, \dots, n]$, compute $\sum_{i = 1}^{n} v_i$.

    \item \verb|sort|: Sort the elements of $v$, where $v$ is an array of $n$ elements randomly shuffled, such that $v_i \in [1, n]$ and $v_i \neq v_j, \forall i \neq j$.
\end{description}

The full list of algorithms supported by pSTL-Bench\xspace can be found in 
Table\xspace~\ref{tab:available_benchmarks} in Appendix\xspace~\ref{app:table_pstl_algs}.

Since each implementation might have a slightly different interface, the calls to the algorithms are wrapped within lambda functions, as shown in Listings\xspace~\ref{lst:stl-for-each-wrapper} and~\ref{lst:gnu-for-each-wrapper}.
These lambda functions are then called from a helper function that executes them repeatedly to measure their execution time with Google's Benchmark library~\cite{google_benchmark}, see Listing\xspace~\ref{lst:for-each-wrapper}. The wrapping function simply measures the time taken for the invocation of the function to be executed (function~\verb|f|).

This structure facilitates the incorporation of new benchmarks while the compiler can still inline the functions to avoid the potential overhead of calling the lambdas.

It should be mentioned that it is possible to manually select the input size and data types (\verb|int|, \verb|float|, \verb|double|, etc.) used in the benchmarks, therefore, extending its customization and analysis capabilities.

\begin{lstlisting}[float, floatplacement=thbp, breaklines=true, caption={Wrapper of the \texttt{for\_each} benchmark using STL.}, label={lst:stl-for-each-wrapper}]
const auto for_each_std = [](auto && policy, const auto & input_data, auto && f) {
    std::for_each(policy, input_data.begin(), 
                  input_data.end(), f);
};
\end{lstlisting}

\begin{lstlisting}[float, floatplacement=thbp, breaklines=true, caption={Wrapper of the \texttt{for\_each} benchmark using GNU's implementation.}, label={lst:gnu-for-each-wrapper}]
const auto sort_gnu = []([[maybe_unused]] auto && executionPolicy, auto & input_data) {
    __gnu_parallel::sort(input_data.begin(), 
                         input_data.end());
};
\end{lstlisting}

\begin{lstlisting}[float, floatplacement=thbp, breaklines=true, caption={Example of the benchmark wrapper for \texttt{for\_each}.}, label={lst:for-each-wrapper}]
template<class Policy, class Function>
static void benchmark_for_each_wrapper(benchmark::State & state, Function && f) {
    constexpr auto execution_policy = Policy{};
    const auto & size = state.range(0);
    auto data = suite::generate_increment<Policy>(execution_policy, size, 1);

    for (auto _ : state) {
        WRAP_TIMING(f(execution_policy, data, kernel);)
    }

    state.SetBytesProcessed(state.iterations() * data.size() * sizeof(int));
}
\end{lstlisting}

\section{Experimental Evaluation}
\label{sec:exp_eval}
For acquiring relevant results, experiments are carried out using different configurations in terms of hardware, software, number of threads, input sizes, and memory allocation.

\subsection{Hardware and Software Setup}

We conduct experiments on three different multi-core, shared-memory
systems that comprise \num{32}, \num{64}, and \num{128} cores,
which are called \emph{Hydra}\xspace, \emph{Nebula}\xspace, and \mbox{\emph{VSC-5}}\xspace,
respectively.\footnote{The names of the machines were intentionally anonymized for the dual-anonymous review process.}
On these systems, we evaluate three different compilers: GCC, Intel oneAPI compiler, and NVIDIA\xspace HPC SDK.
Also, several backends are tested, which are: Intel's Threading Building Blocks (TBB), High-Performance ParalleX (HPX), and OpenMP\xspace through the implementations of GNU and NVIDIA\xspace.

Furthermore, we also perform experiments on two GPU-based systems, \mbox{\emph{Tesla}}\xspace and \mbox{\emph{Ampere}}\xspace, using the NVIDIA\xspace HPC SDK with the CUDA compiler.

We provide an overview of the hardware and software details in
Table\xspace~\ref{tab:machines}.
\begin{table*}[tb]
  \caption{Summary of the hardware and software used in our study.}
  \label{tab:machines}
  \centering
  \setlength{\tabcolsep}{4pt}
\begin{small}
  \begin{tabular}{lrrrrr}
  \toprule
  machine & \emph{Hydra}\xspace & \emph{Nebula}\xspace & \mbox{\emph{VSC-5}}\xspace & \mbox{\emph{Tesla}}\xspace & \mbox{\emph{Ampere}}\xspace \\
  \midrule
  CPU/GPU & Intel Xeon 6130F & AMD EPYC 7551 & AMD EPYC 7713 & NVIDIA\xspace Tesla P4 & NVIDIA\xspace Ampere A2 \\
  core frequency & \qty{2.10}{\giga\hertz} & \qty{2.00}{\giga\hertz} & \qty{2.00}{\giga\hertz} & \qty{1.11}{\giga\hertz} & \qty{1.77}{\giga\hertz} \\
  \#sockets & 2 & 2 & 2 & --- & --- \\
  \#cores (node) & 32 & 64 & 128 & 2560 & 1280 \\
  \#threads (node) & 32 & 64 & 256 & 2560 & 1280 \\
  compilers & g++~12.1.0\xspace & g++~12.3.0\xspace & g++~12.2.0\xspace & g++~10.2.1\xspace & g++~10.2.1\xspace \\
            & nvc++~22.11\xspace & nvc++~23.7\xspace & nvc++~22.9\xspace & nvc++~23.5\xspace & nvc++~23.5\xspace \\
            & icpx~2021.7.0\xspace &  & icpx~2022.2.1\xspace & & \\
  libraries & TBB~2021.9.0\xspace & TBB~2021.10.0\xspace & TBB~2021.7.0\xspace & CUDA~11.8\xspace & CUDA~12.2\xspace \\
            & HPX~1.9.0\xspace & HPX~1.9.1\xspace & HPX~1.8.1\xspace & & \\
            & GOMP~(from g++~12.1.0\xspace)\xspace & GOMP~(from g++~12.3.0\xspace)\xspace & GOMP~(from g++~12.2.0\xspace)\xspace & & \\
            & NVOMP~22.11\xspace & NVOMP~22.11\xspace & NVOMP~22.9\xspace & & \\
  \bottomrule
  \end{tabular}
\end{small}
\end{table*}

\subsection{Experimental Setup}

In our experiments we test the parallel algorithms with different numbers of threads and problem sizes.
Specifically, the thread counts range from 1 to the maximum available core count on each machine, so the values $1, 2, 4, \dots, \#\text{cores}$ are used.
Concerning problem sizes, inputs vary from \num[parse-numbers=false]{2^3} to \num[parse-numbers=false]{2^{30}} elements, using 32-bit integers, enabling the testing of a broad set of inputs that are accommodated in various cache levels or necessitate consistent DRAM traffic.

\subsection{The Impact of Memory Allocation}
Memory allocation might have a significant impact on performance in modern computer architectures.
For that reason, a custom parallel allocator is used for the experiments.
Through the first-touch policy of Linux, this allocator makes each thread touch the first byte of each object with the given parallel policy, as shown in Listing\xspace~\ref{lst:parallel-allocator}.
This enforces the correct placement of threads and memory pages.
The parallel allocator used in this work is an adapted version of the NUMA allocator that is part of HPX.\footnote{
For the NUMA allocator of HPX, see
\url{https://github.com/STEllAR-GROUP/hpx/blob/fe048ee6e01abedad0a60a0fdc204116419871c3/libs/core/compute_local/include/hpx/compute_local/host/numa_allocator.hpp}
}

\begin{lstlisting}[float, breaklines=true, caption={Function \texttt{allocate} for the custom parallel allocator.}, label={lst:parallel-allocator}]
pointer allocate(size_type cnt, void const * = nullptr) {
    // allocate memory
    auto p = static_cast<pointer>(::operator new(cnt * sizeof(T)));

    // touch the first byte of every object
    std::for_each(execution_policy, begin, begin + cnt, [](T & val) { *reinterpret_cast<char *>(&val) = 0; });

    // return the overall memory block
    return p;
}
\end{lstlisting}

\begin{figure}[t]
    \centering
    \includegraphics[width=\linewidth]{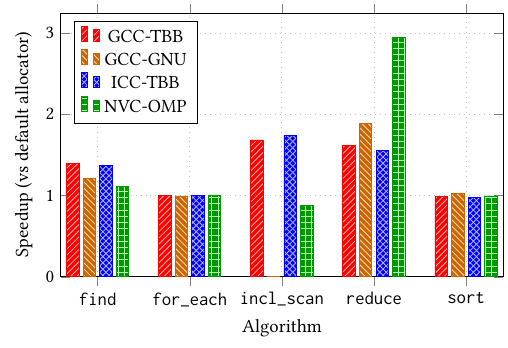}
    \caption{Speedup when using custom parallel allocator with 32 threads and a problem size of \num[parse-numbers=false]{2^{30}} elements in \emph{Hydra}\xspace. Higher is better\xspace.}
    \label{fig:speedup-parallel-allocator}
\end{figure}

Fig.\xspace~\ref{fig:speedup-parallel-allocator} shows the speedup achieved when using the custom parallel allocator in \emph{Hydra}\xspace.
Using this allocator leads to an increase in performance in three of the five algorithms.
For the \texttt{X::find}\xspace algorithm, the improvement goes from \qtyrange{11}{39}{\percent}, from \qtyrange{-13}{74}{\percent} for the \texttt{X::inclusive\_scan}\xspace and \qtyrange{56}{194}{\percent} for \texttt{X::reduce}\xspace.
In the \texttt{X::for\_each}\xspace and the \texttt{X::sort}\xspace algorithms, the differences are smaller, always bellow \qty{2.5}{\percent} and typically around \qty{0.5}{\percent}.

In only one experiment (\texttt{X::inclusive\_scan}\xspace), the parallel allocator notably reduced performance, that is with the NVIDIA\xspace compiler and the OpenMP\xspace backend.

Considering the results of the parallel allocator, the rest of the benchmarks are executed using it except for those tests related to HPX, which is incompatible with it.

\section{Experimental Results} \label{sec:experimental-results}

Table\xspace~\ref{tab:speedup-vs-seq} summarises the results obtained, showing the speedup compared to the sequential implementation with GCC. In this table, we use a fixed baseline performance, GCC's sequential performance in this case, which may lead to speedups that are large than the total core count.

Table\xspace~\ref{tab:threads-with-good-efficiency} shows the maximum number of threads for which parallel efficiency is above \qty{70}{\percent}.
The purpose of this table is to analyze the number of threads that can be effectively utilized without excessively wasting resources.
The threshold of \qty{70}{\percent} is somewhat arbitrarily defined, but it assists in estimating the effectiveness of the parallelization.

All the data presented in this section are derived from the median of ten executions.

\begin{table*}[t]
\caption{Speedup against GCC's sequential implementation for machines \emph{Hydra}\xspace, \emph{Nebula}\xspace, and \mbox{\emph{VSC-5}}\xspace, with \num{32}, \num{64}, and \num{128} cores, respectively. Notation is \emph{Hydra}\xspace/\emph{Nebula}\xspace/\mbox{\emph{VSC-5}}\xspace. Higher is better\xspace.}
\label{tab:speedup-vs-seq}
\begin{tabular}{rlllll}
\toprule
 & \texttt{X::find}\xspace & \texttt{X::for\_each}\xspace & \texttt{X::inclusive\_scan}\xspace & \texttt{X::reduce}\xspace & \texttt{X::sort}\xspace \\
\midrule
GCC-TBB & 9.54 / 3.99 / 5.83 & 31.92 / 54.31 / 85.18 & 4.72 / 3.31 / 4.80 & 12.29 / 6.25 / 7.14 & 9.51 / 10.04 / 9.42 \\
GCC-GNU & 6.72 / 1.46 / 2.65 & 31.68 / 54.15 / 85.01 & --- / --- / --- & 15.08 / 5.49 / 7.20 & 28.09 / 28.07 / 65.56 \\
GCC-HPX & 7.30 / 1.07 / 2.31 & 29.17 / 50.52 / 64.33 & 2.18 / 0.79 / 0.84 & 9.46 / 1.14 / 1.76 & 10.24 / 9.10 / 8.46 \\
ICC-TBB & 10.02 / --- / 5.77 & 56.53 / --- / 154.08 & 4.68 / --- / 4.81 & 12.37 / --- / 7.19 & 10.05 / --- / 9.78 \\
NVC-OMP & 4.78 / 1.07 / 1.31 & 224.55 / 305.65 / 608.00 & 0.88 / 0.62 / 0.96 & 23.77 / 17.91 / 11.54 & 9.14 / 7.45 / 6.76 \\
\bottomrule
\end{tabular}
\end{table*}

\begin{table*}[t]
\caption{Maximum number of threads for which parallel efficiency is above \qty{70}{\percent} (compared to the execution with \num{1} thread) for machines \emph{Hydra}\xspace, \emph{Nebula}\xspace, and \mbox{\emph{VSC-5}}\xspace. Notation is \emph{Hydra}\xspace/\emph{Nebula}\xspace/\mbox{\emph{VSC-5}}\xspace. Higher is better\xspace.}
\label{tab:threads-with-good-efficiency}
\begin{tabular}{rlllll}
\toprule
 & \texttt{X::find}\xspace & \texttt{X::for\_each}\xspace & \texttt{X::inclusive\_scan}\xspace & \texttt{X::reduce}\xspace & \texttt{X::sort}\xspace \\
\midrule
GCC-TBB & 1 / 2 / 1 & 32 / 64 / 32 & 1 / 1 / 1 & 8 / 4 / 1 & 4 / 8 / 4 \\
GCC-GNU & 1 / 1 / 1 & 32 / 64 / 32 & --- / --- / --- & 1 / 4 / 1 & 32 / 32 / 16 \\
GCC-HPX & 16 / 4 / 1 & 32 / 64 / 8 & 4 / 2 / 1 & 8 / 2 / 1 & 8 / 4 / 4 \\
ICC-TBB & 1 / --- / 1 & 32 / --- / 16 & 1 / --- / 1 & 8 / --- / 1 & 8 / --- / 4 \\
NVC-OMP & 8 / 4 / 1 & 32 / 32 / 128 & 1 / 1 / 1 & 1 / 2 / 1 & 2 / 2 / 4 \\
\bottomrule
\end{tabular}
\end{table*}

\subsection{\texttt{X::find}\xspace}
\begin{figure}[t]
    \centering
    \includegraphics[width=\linewidth]{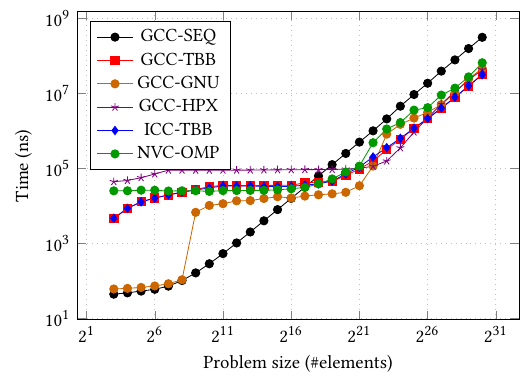}
    \caption{Execution time scalability with the problem size in \emph{Hydra}\xspace. All cores were used except for GCC’s sequential implementation. Lower is better\xspace.}
    \label{fig:hydra-find-problem-size}
\end{figure}

Figure\xspace~\ref{fig:hydra-find-problem-size} shows how the execution times grow with the size of the problem in \emph{Hydra}\xspace.
The behavior for other machines is shown in Appendix\xspace~\ref{ap:find}.
For small problem sizes, the sequential implementation offers better execution times, with a difference of orders of magnitude, highlighting the overhead of launching parallel sections.
This difference is consistently reduced until problem sizes of around \num[parse-numbers=false]{2^{16}}, where the performance of the parallelism starts to compensate for the overhead.
It is interesting to note that the GNU parallel implementation falls back to the sequential implementation until the problem size reaches \num[parse-numbers=false]{2^9}.
From problem sizes of \num[parse-numbers=false]{2^{20}}, parallel implementations begin to be significantly faster than the sequential one.
These behaviors, regarding the input size, can be extended to the rest of the benchmarks.

Related to the speedup, efficiency is an issue, where the maximum speedup achieved was $10\times$ with the Intel compiler and using TBB with 32 threads in \emph{Hydra}\xspace.
Also, using more than 32 threads results in little, if any, gains in performance.

\subsection{\texttt{X::for\_each}\xspace}
\begin{figure*}[t]
    \centering
    \begin{subfigure}{0.33\textwidth}
        \centering
        \includegraphics[width=\textwidth]{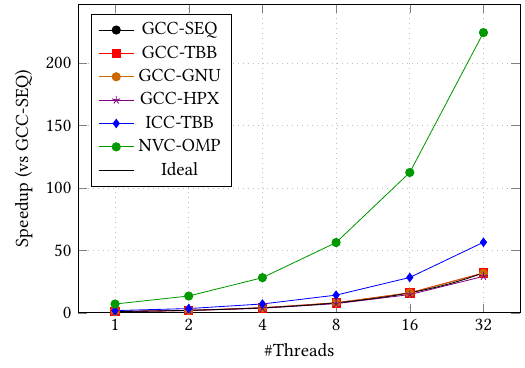}
        \caption{\emph{Hydra}\xspace}
        \label{fig:hydra-foreach-speedup}
    \end{subfigure}
    \hfill
    \begin{subfigure}{0.33\textwidth}
        \centering
        \includegraphics[width=\textwidth]{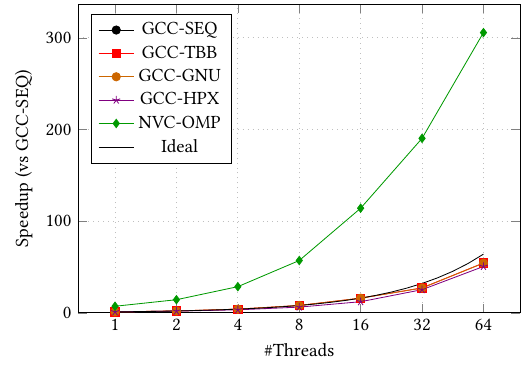}
        \caption{\emph{Nebula}\xspace}
        \label{fig:nebula-foreach-speedup}
    \end{subfigure}
    \hfill
    \begin{subfigure}{0.33\textwidth}
        \centering
        \includegraphics[width=\textwidth]{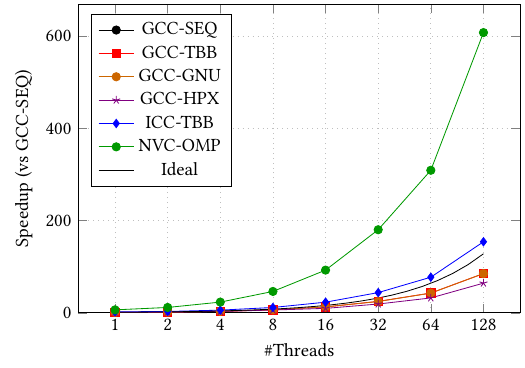}
        \caption{\mbox{\emph{VSC-5}}\xspace}
        \label{fig:vsc-foreach-speedup}
    \end{subfigure}
    \caption{Speedup against GCC's sequential implementation for benchmark \texttt{X::for\_each}\xspace. Higher is better\xspace.}
    \label{fig:foreach-speedup}
\end{figure*}

Figure\xspace~\ref{fig:foreach-speedup} shows the speedup obtained for a problem size of \num[parse-numbers=false]{2^{30}} elements.
Note that the Intel compiler and, particularly, the NVIDIA\xspace compiler achieve a better performance than GCC.
Profiling of the code indicates that besides using its math library, the NVIDIA\xspace and Intel compilers make better use of vector instructions in comparison to GNU's.
Interestingly, the NVIDIA\xspace compiler uses \num{256}-bit instructions, while the Intel compiler uses only \num{128}-bit instructions.

Additionally, data in Table\xspace~\ref{tab:threads-with-good-efficiency} shows an almost perfect scaling.
This is reasonable given the \textit{embarrassingly parallel} nature of the algorithm.

\subsection{\texttt{X::inclusive\_scan}\xspace}
First, it should be noted that the GNU's collection of parallel algorithms does not include the \verb|inclusive_scan| function.

\begin{figure*}[t]
    \centering
    \begin{subfigure}{0.33\textwidth}
        \centering
        \includegraphics[width=\textwidth]{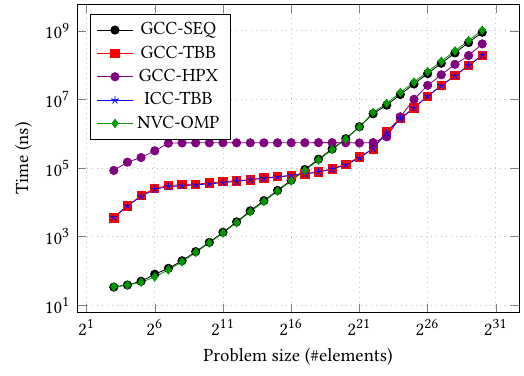}
        \caption{\emph{Hydra}\xspace}
        \label{fig:hydra-inclusive-scan-problem-size}
    \end{subfigure}
    \hfill
    \begin{subfigure}{0.33\textwidth}
        \centering
        \includegraphics[width=\textwidth]{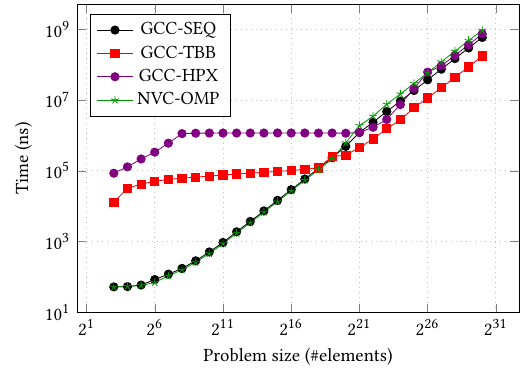}
        \caption{\emph{Nebula}\xspace}
        \label{fig:nebula-inclusive-scan-problem-size}
    \end{subfigure}
    \hfill
    \begin{subfigure}{0.33\textwidth}
        \centering
        \includegraphics[width=\textwidth]{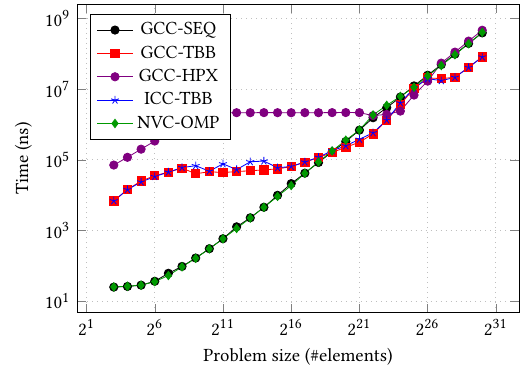}
        \caption{\mbox{\emph{VSC-5}}\xspace}
        \label{fig:vsc-inclusive-scan-problem-size}
    \end{subfigure}
    \caption{Execution time scalability with the problem size. All cores used except for GCC’s seq. implementation\xspace. Benchmark \texttt{X::inclusive\_scan}\xspace. Lower is better\xspace.}
    \label{fig:inclusive-scan-problem-size}
\end{figure*}

\begin{figure*}[t]
    \centering
    \begin{subfigure}{0.33\textwidth}
        \centering
        \includegraphics[width=\textwidth]{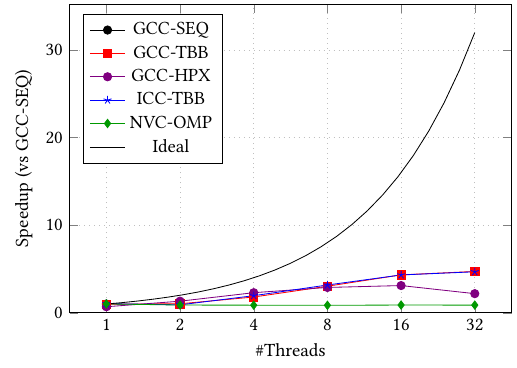}
        \caption{\emph{Hydra}\xspace}
        \label{fig:hydra-inclusive-scan-speedup}
    \end{subfigure}
    \hfill
    \begin{subfigure}{0.33\textwidth}
        \centering
        \includegraphics[width=\textwidth]{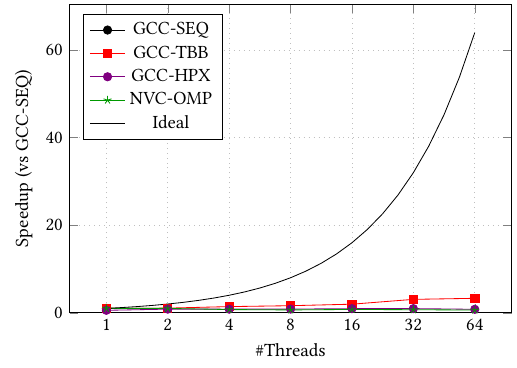}
        \caption{\emph{Nebula}\xspace}
        \label{fig:nebula-inclusive-scan-speedup}
    \end{subfigure}
    \hfill
    \begin{subfigure}{0.33\textwidth}
        \centering
        \includegraphics[width=\textwidth]{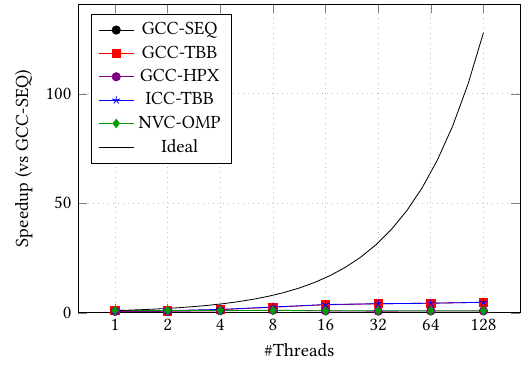}
        \caption{\mbox{\emph{VSC-5}}\xspace}
        \label{fig:vsc-inclusive-scan-speedup}
    \end{subfigure}
    \caption{Speedup against GCC's seq. implementation. Problem size is $2^{30}$\xspace. Benchmark \texttt{X::inclusive\_scan}\xspace. Higher is better\xspace.}
    \label{fig:inclusive-scan-speedup}
\end{figure*}

Based on the results shown in Figures\xspace~\ref{fig:inclusive-scan-problem-size}, \ref{fig:inclusive-scan-speedup}, and Table\xspace~\ref{tab:speedup-vs-seq}, it is evident that the NVIDIA\xspace implementation yields limited speedup for this operation, being even slower than the sequential version.
This is because this implementation falls back to a sequential execution, which is negatively affected by the custom parallel allocator used.

The rest of the implementations produce slight improvements in performance when using parallelism.
For example, a maximum speedup of $4.81\times$ can be achieved using TBB as the backend.
However, the efficiency of these implementations is far from ideal.

\subsection{\texttt{X::reduce}\xspace}
It should be noted that the GNU's collection of algorithms does not include a \verb|reduction| function so the \verb|accumulate| function has been used instead.

\begin{figure*}[t]
    \centering
    \begin{subfigure}{0.33\textwidth}
        \centering
        \includegraphics[width=\textwidth]{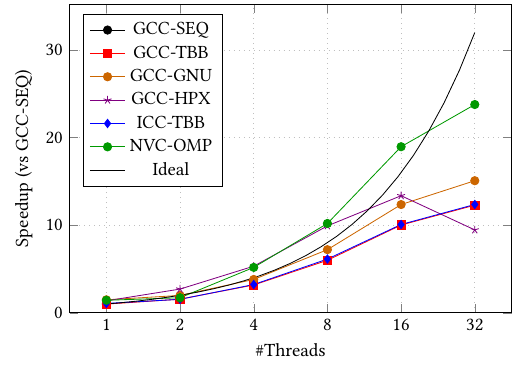}
        \caption{\emph{Hydra}\xspace}
        \label{fig:reduce-speedup-hydra}
    \end{subfigure}
    \hfill
    \begin{subfigure}{0.33\textwidth}
        \centering
        \includegraphics[width=\textwidth]{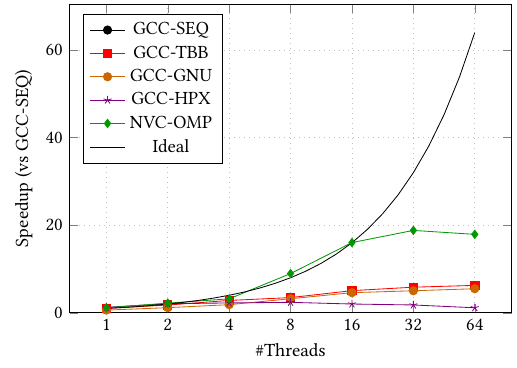}
        \caption{\emph{Nebula}\xspace}
        \label{fig:reduce-speedup-nebula}
    \end{subfigure}
    \hfill
    \begin{subfigure}{0.33\textwidth}
        \centering
        \includegraphics[width=\textwidth]{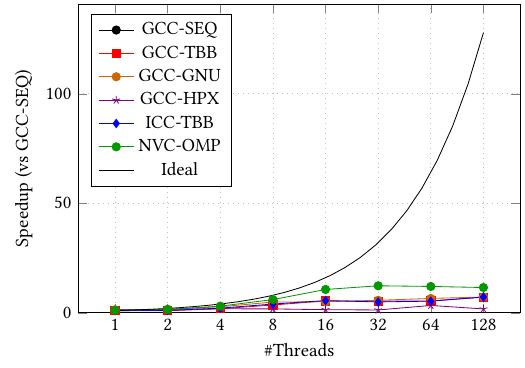}
        \caption{\mbox{\emph{VSC-5}}\xspace}
        \label{fig:reduce-speedup-vsc}
    \end{subfigure}
    \caption{Speedup against GCC's seq. implementation. Problem size is $2^{30}$\xspace. Benchmark \texttt{X::reduce}\xspace. Higher is better\xspace.}
    \label{fig:reduce-speedup}
\end{figure*}

As shown in Figure\xspace~\ref{fig:reduce-speedup} the speedup behavior changes from system to system.
In \emph{Hydra}\xspace (Figure\xspace~\ref{fig:reduce-speedup-hydra}), for example, a good speedup is achieved up to \num{16} threads, with a maximum speedup of $23.77\times$.

However, the performance achieved with the NVIDIA\xspace compiler with the OpenMP backend stands out, especially in \emph{Nebula}\xspace and \mbox{\emph{VSC-5}}\xspace.
In these systems, the rest of the compilers and backends present poor scalability, especially when more than \num{16} threads are used, see Figures\xspace~\ref{fig:reduce-speedup-nebula} and~\ref{fig:reduce-speedup-vsc}.

\subsection{\texttt{X::sort}\xspace}
Two different factors are relevant when analyzing the results for this benchmark, the problem size and the number of threads.

\begin{figure*}
    \centering
    \begin{subfigure}{0.33\textwidth}
        \centering
        \includegraphics[width=\textwidth]{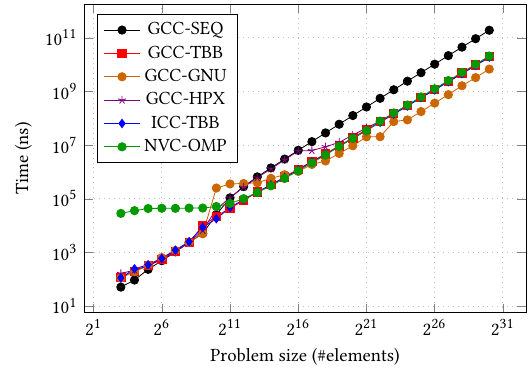}
        \caption{\emph{Hydra}\xspace}
        \label{fig:sort-problem-size-hydra}
    \end{subfigure}
    \hfill
    \begin{subfigure}{0.33\textwidth}
        \centering
        \includegraphics[width=\textwidth]{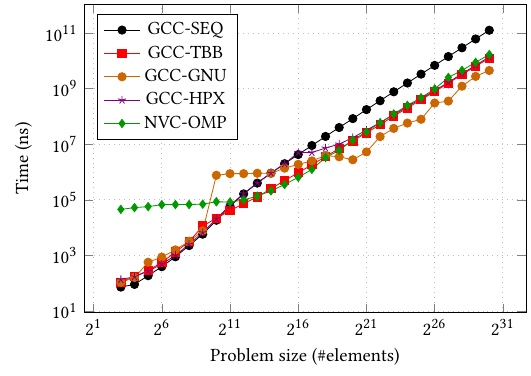}
        \caption{\emph{Nebula}\xspace}
        \label{fig:sort-problem-size-nebula}
    \end{subfigure}
    \hfill
    \begin{subfigure}{0.33\textwidth}
        \centering
        \includegraphics[width=\textwidth]{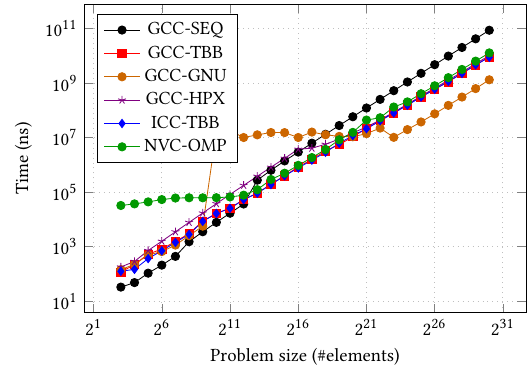}
        \caption{\mbox{\emph{VSC-5}}\xspace}
        \label{fig:sort-problem-size-vsc}
    \end{subfigure}
    \caption{Execution time scalability with the problem size. All cores used except for GCC’s seq. implementation\xspace. Benchmark \texttt{X::sort}\xspace. Lower is better\xspace.}
    \label{fig:sort-problem-size}
\end{figure*}

For small sizes (see Figure\xspace~\ref{fig:sort-problem-size}), below \num[parse-numbers=false]{2^{12}}, the NVC-OMP shows a significantly higher overhead than the rest of compilers and backends.
For bigger sizes, it is the GNU implementation that obtains better results.

\begin{figure*}
    \centering
    \begin{subfigure}{0.33\textwidth}
        \centering
        \includegraphics[width=\textwidth]{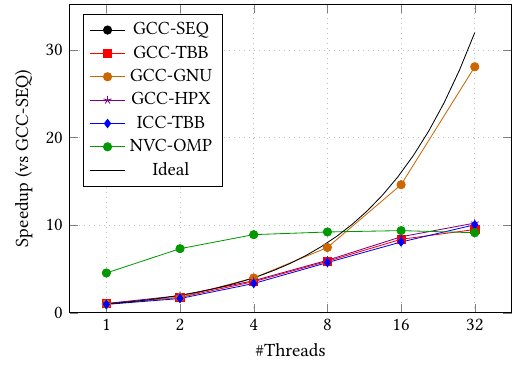}
        \caption{\emph{Hydra}\xspace}
        \label{fig:sort-speedup-hydra}
    \end{subfigure}
    \hfill
    \begin{subfigure}{0.33\textwidth}
        \centering
        \includegraphics[width=\textwidth]{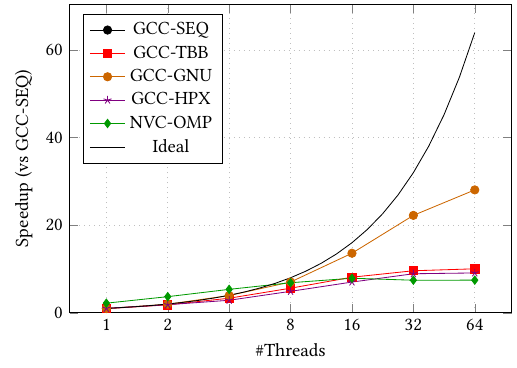}
        \caption{\emph{Nebula}\xspace}
        \label{fig:sort-speedup-nebula}
    \end{subfigure}
    \hfill
    \begin{subfigure}{0.33\textwidth}
        \centering
        \includegraphics[width=\textwidth]{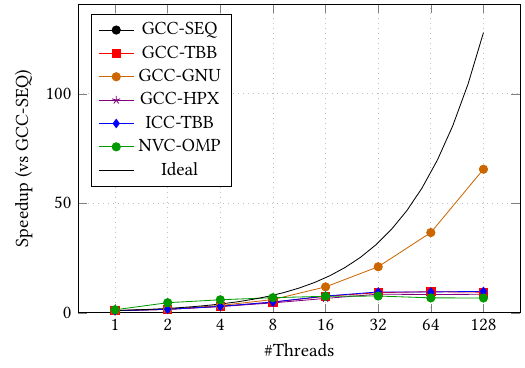}
        \caption{\mbox{\emph{VSC-5}}\xspace}
        \label{fig:sort-speedup-vsc}
    \end{subfigure}
    \caption{Speedup against GCC's seq. implementation. Problem size is $2^{30}$\xspace. Benchmark \texttt{X::sort}\xspace. Higher is better\xspace.}
    \label{fig:sort-speedup}
\end{figure*}

Regarding the number of threads, when this is \num{8} or less, the NVIDIA\xspace compiler achieves a considerably better performance than the rest, see Figure\xspace~\ref{fig:sort-speedup}.
Nevertheless, with more threads, the implementation of GNU which uses OpenMP\xspace establishes itself as a clear winner.

It is interesting to note that GNU's efficiency is above \qty{70}{\percent} up to \num{32} threads in \emph{Hydra}\xspace and \emph{Nebula}\xspace, but only up to \num{16} threads in \mbox{\emph{VSC-5}}\xspace, see Table\xspace~\ref{tab:threads-with-good-efficiency}.
The rest of the backends, cannot use more than \num{8} threads while keeping an efficiency of \qty{70}{\percent} or better.

\subsection{Performance on GPUs}
\begin{figure*}
    \centering
    \begin{subfigure}{0.33\textwidth}
        \centering
        \includegraphics[width=\textwidth]{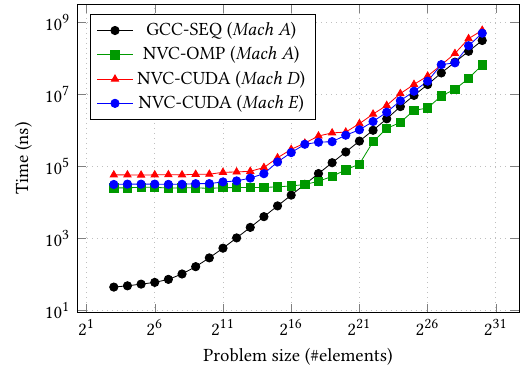}
        \caption{\texttt{X::find}\xspace}
        \label{fig:find-problem-size-hydra-gpu}
    \end{subfigure}
    \hfill
    \begin{subfigure}{0.33\textwidth}
        \centering
        \includegraphics[width=\textwidth]{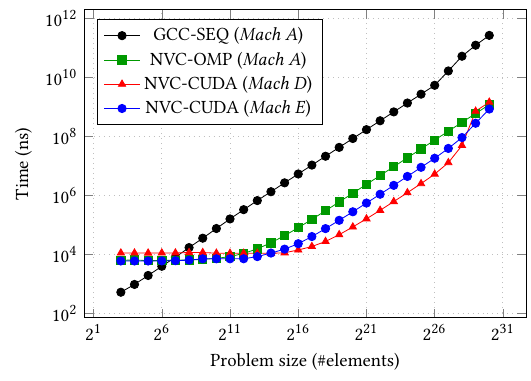}
        \caption{\texttt{X::for\_each}\xspace}
        \label{fig:for-each-problem-size-hydra-gpu}
    \end{subfigure}
    \hfill
    \begin{subfigure}{0.33\textwidth}
        \centering
        \includegraphics[width=\textwidth]{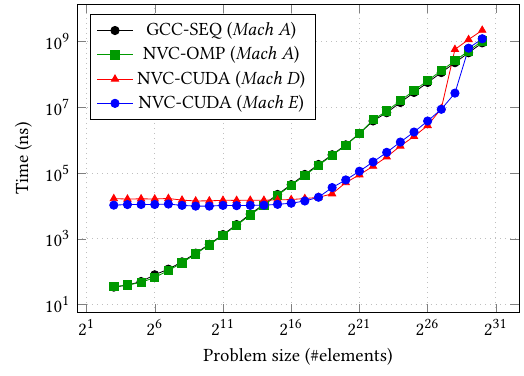}
        \caption{\texttt{X::inclusive\_scan}\xspace}
        \label{fig:incl-scan-problem-size-hydra-gpu}
    \end{subfigure}
    \hfill
    \begin{subfigure}{0.33\textwidth}
        \centering
        \includegraphics[width=\textwidth]{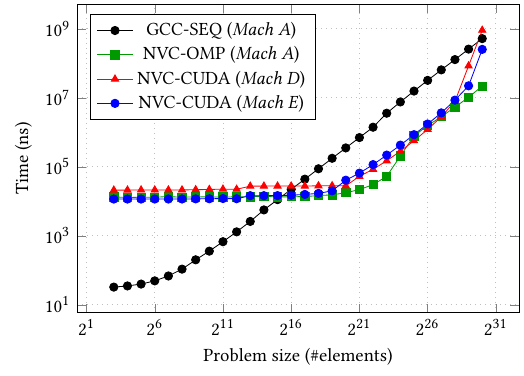}
        \caption{\texttt{X::reduce}\xspace}
        \label{fig:reduce-problem-size-hydra-gpu}
    \end{subfigure}
    \begin{subfigure}{0.33\textwidth}
        \centering
        \includegraphics[width=\textwidth]{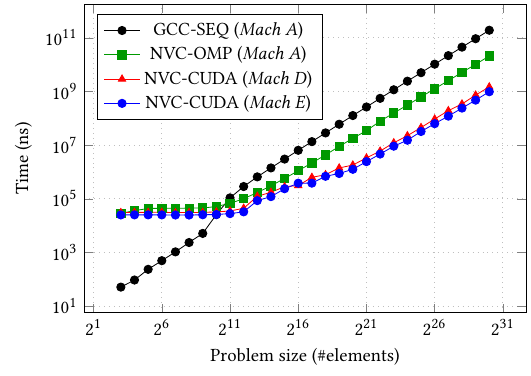}
        \caption{\texttt{X::sort}\xspace}
        \label{fig:sort-problem-size-hydra-gpu}
    \end{subfigure}
    \caption{Execution time scalability with the problem size. All cores used except for GCC’s seq. implementation\xspace. Lower is better\xspace.}
    \label{fig:problem-size-hydra-gpu}
\end{figure*}

Figure\xspace~\ref{fig:problem-size-hydra-gpu} includes the execution times of the benchmarks on \emph{Hydra}\xspace, \mbox{\emph{Tesla}}\xspace and \mbox{\emph{Ampere}}\xspace.
Results show that both the input size and the particular algorithm play a key role in terms of performance for the GPUs compared to the CPU.

First, a substantial penalty is paid in the form of overhead when using a small input size.
Also, transferring a large amount of data seems to penalize the performance of the GPUs as well, see Figures\xspace~\ref{fig:incl-scan-problem-size-hydra-gpu} and~\ref{fig:reduce-problem-size-hydra-gpu}.

Regarding the algorithms, the \texttt{X::find}\xspace and \texttt{X::reduce}\xspace have poor performance on the GPUs.
However, the \texttt{X::for\_each}\xspace, the \texttt{X::inclusive\_scan}\xspace, and the \texttt{X::sort}\xspace achieve significantly better execution times, especially for the sizes ranging from \num[parse-numbers=false]{2^{16}} to \num[parse-numbers=false]{2^{27}} elements.

\section{Conclusions}
\label{sec:conclusions}

In this work, we introduced a specialized set of benchmarks named pSTL-Bench\xspace to assess the performance of the parallel algorithms present in the C$++$\xspace standard template library.
With this suite, we aim to facilitate the users comparing compilers and backends to decide which fits best the particular characteristics of their system.

Additionally, using this set of benchmarks, we conducted a comprehensive analysis of performance on current multi-core architectures by benchmarking some of the most used algorithms in STL.
This analysis took into account not only the compiler and backend being used but also the input size, the number of threads used, and the memory allocation.

First, we showed the potential performance improvements when using a custom parallel allocator in NUMA systems.
Generally, execution times are improved, up to \qty{194}{\percent} in the best case, and minor losses in some scenarios.

Second, when using small input sizes it is possible to note the large overhead of launching parallel sections and the differences between backends, being the TBB typically lower than those based on OpenMP\xspace or HPX.
It is interesting to note as well that the GNU implementation falls back to the sequential execution for inputs lower than \num{512} elements.

On the other hand, for large input sizes, the parallel implementations show better in the vast majority of scenarios.
For example, the speedup and efficiency in the \texttt{X::for\_each}\xspace stand out compared to the rest of the algorithms.
Nevertheless, some algorithms like \texttt{X::find}\xspace or \texttt{X::inclusive\_scan}\xspace show poor speedup and, consequently, a bad efficiency.

Third, it is interesting to note that there might be large differences in performance due to vectorization.
In our experiments, we found that the NVIDIA\xspace compiler applies vector instruction more often than the rest of the compilers, particularly in the \texttt{X::for\_each}\xspace algorithm.

Finally, we evaluated the performance of the CUDA backend for the NVIDIA\xspace compiler with two different GPUs.
We found high penalties for small inputs, but also for high inputs where the data transfers dominate the execution times.
Nevertheless, for medium problem sizes and some particular algorithms, the GPUs are capable of outperforming the CPUs.

Regarding future work, we would like to expand our benchmark suite, to include more algorithms 
and also to support more compilers and backends.
Similarly, it would be interesting to extend our experimental analysis on more architectures, particularly GPUs and, potentially, FPGAs.

\FloatBarrier
\bibliographystyle{ACM-Reference-Format}
\bibliography{main-clean}\balance

\newpage
\FloatBarrier
\clearpage
\appendix
\onecolumn

\clearpage
\section{Support of Parallel Implementations}
\label{app:table_pstl_algs}
\begin{table*}[ht]
\caption{List of algorithms with support for parallelization in STL. Those with benchmarks in pSTL-Bench are noted in gray.}
\label{tab:available_benchmarks}
\centering
\begin{tabular}{l l l}
\toprule
\cellcolor{gray!15}\verb|adjacent_difference| & \cellcolor{gray!15}\verb|adjacent_find| & \cellcolor{gray!15}\verb|all_of| \\
\cellcolor{gray!15}\verb|any_of| & \cellcolor{gray!15}\verb|copy| & \cellcolor{gray!15}\verb|copy_if| \\
\cellcolor{gray!15}\verb|copy_n| & \cellcolor{gray!15}\verb|count| & \cellcolor{gray!15}\verb|count_if| \\
\verb|destroy| & \verb|destroy_n| & \cellcolor{gray!15}\verb|equal| \\
\cellcolor{gray!15}\verb|exclusive_scan| & \cellcolor{gray!15}\verb|fill| & \verb|fill_n| \\
\cellcolor{gray!15}\verb|find| & \verb|find_end| & \verb|find_first_of| \\
\verb|find_if| & \verb|find_if_not| & \cellcolor{gray!15}\verb|for_each| \\
\verb|for_each_n| & \cellcolor{gray!15}\verb|generate| & \verb|generate_n| \\
\cellcolor{gray!15}\verb|includes| & \cellcolor{gray!15}\verb|inclusive_scan| & \cellcolor{gray!15}\verb|inplace_merge| \\
\verb|is_heap| & \verb|is_heap_until| & \verb|is_partitioned| \\
\cellcolor{gray!15}\verb|is_sorted| & \verb|is_sorted_until| & \cellcolor{gray!15}\verb|lexicographical_compare| \\
\cellcolor{gray!15}\verb|max_element| & \cellcolor{gray!15}\verb|merge| & \cellcolor{gray!15}\verb|min_element| \\
\verb|minmax_element| & \cellcolor{gray!15}\verb|mismatch| & \verb|move| \\
\cellcolor{gray!15}\verb|none_of| & \verb|nth_element| & \cellcolor{gray!15}\verb|partial_sort| \\
\verb|partial_sort_copy| & \cellcolor{gray!15}\verb|partition| & \verb|partition_copy| \\
\cellcolor{gray!15}\verb|reduce| & \verb|remove| & \verb|remove_copy| \\
\verb|remove_copy_if| & \verb|remove_if| & \verb|replace| \\
\verb|replace_copy| & \verb|replace_copy_if| & \verb|replace_if| \\
\verb|reverse| & \verb|reverse_copy| & \verb|rotate| \\
\verb|rotate_copy| & \cellcolor{gray!15}\verb|search| & \verb|search_n| \\
\cellcolor{gray!15}\verb|set_difference| & \cellcolor{gray!15}\verb|set_intersection| & \verb|set_symmetric_difference| \\
\verb|set_union| & \cellcolor{gray!15}\verb|sort| & \verb|stable_sort| \\
\verb|stable_partition| & \verb|swap_ranges| & \cellcolor{gray!15}\verb|transform| \\
\cellcolor{gray!15}\verb|transform_exclusive_scan| & \cellcolor{gray!15}\verb|transform_inclusive_scan| & \cellcolor{gray!15}\verb|transform_reduce| \\
\verb|uninitialized_copy| & \verb|uninitialized_copy_n| & \verb|uninitialized_default_construct| \\
\verb|uninitialized_default_construct_n| & \verb|uninitialized_fill| & \verb|uninitialized_fill_n| \\
\verb|uninitialized_move| & \verb|uninitialized_move_n| & \verb|uninitialized_value_construct| \\
\verb|uninitialized_value_construct_n| & \verb|unique| & \verb|unique_copy| \\
\bottomrule
\end{tabular}
\end{table*}

\clearpage
\section{\texttt{X::find}\xspace} \label{ap:find}
\begin{figure*}[h!]
    \centering
    \begin{subfigure}[t]{0.3\textwidth}
        \centering
        \includegraphics[width=\textwidth]{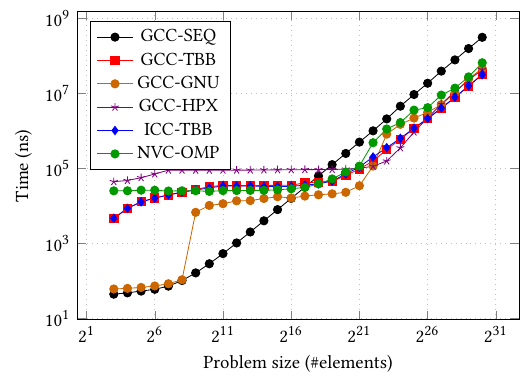}
        \caption{Execution time scalability with the problem size. All cores used except for GCC’s seq. implementation\xspace. Lower is better\xspace.}
    \end{subfigure}
    \hfill
    \begin{subfigure}[t]{0.3\textwidth}
        \centering
        \includegraphics[width=\textwidth]{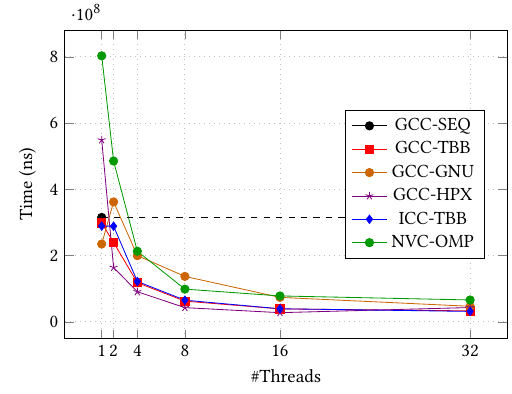}
        \caption{Execution time scalability with the number of threads used. Problem size is $2^{30}$\xspace. Lower is better\xspace.}
    \end{subfigure}
    \hfill
    \begin{subfigure}[t]{0.3\textwidth}
        \centering
        \includegraphics[width=\textwidth]{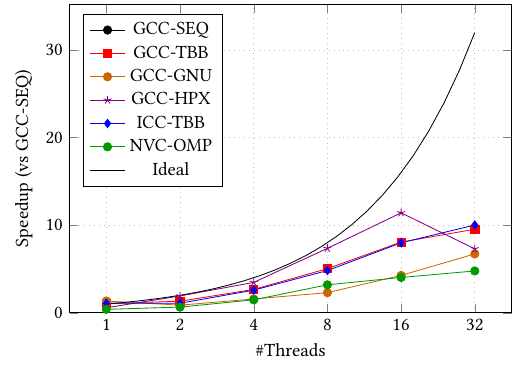}
        \caption{Speedup against GCC's seq. implementation. Problem size is $2^{30}$\xspace. Higher is better\xspace.}
    \end{subfigure}
    \caption{Results for benchmark \texttt{X::find}\xspace; \emph{Hydra}\xspace.}
    \label{fig:hydra-find}
\end{figure*}

\begin{figure*}[h!]
    \centering
    \begin{subfigure}[t]{0.3\textwidth}
        \centering
        \includegraphics[width=\textwidth]{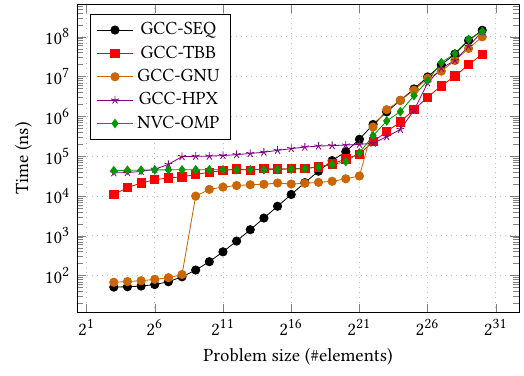}
        \caption{Execution time scalability with the problem size. All cores used except for GCC’s seq. implementation\xspace. Lower is better\xspace.}
    \end{subfigure}
    \hfill
    \begin{subfigure}[t]{0.3\textwidth}
        \centering
        \includegraphics[width=\textwidth]{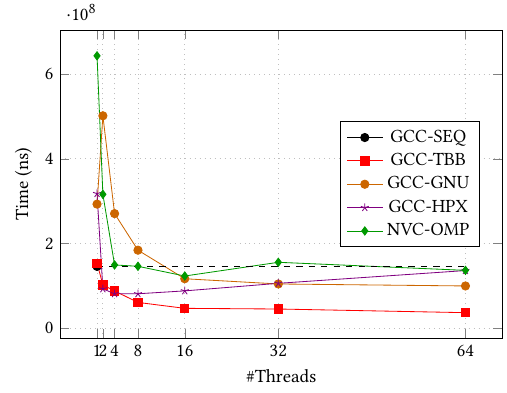}
        \caption{Execution time scalability with the number of threads used. Problem size is $2^{30}$\xspace. Lower is better\xspace.}
    \end{subfigure}
    \hfill
    \begin{subfigure}[t]{0.3\textwidth}
        \centering
        \includegraphics[width=\textwidth]{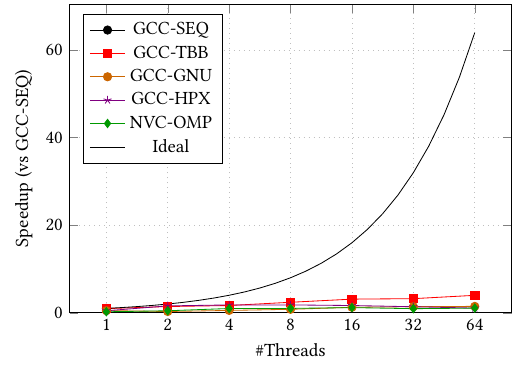}
        \caption{Speedup against GCC's seq. implementation. Problem size is $2^{30}$\xspace. Higher is better\xspace.}
    \end{subfigure}
\end{figure*}

\begin{figure*}[h!]
    \centering
    \begin{subfigure}[t]{0.3\textwidth}
        \centering
        \includegraphics[width=\textwidth]{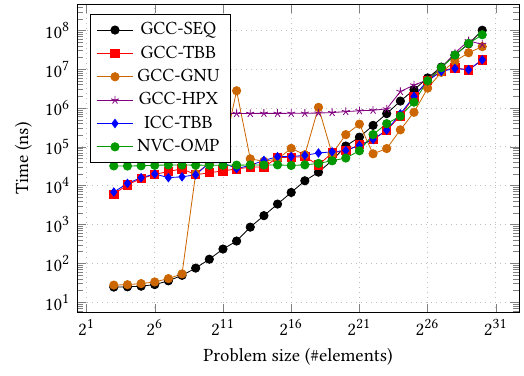}
        \caption{Execution time scalability with the problem size. All cores used except for GCC’s seq. implementation\xspace. Lower is better\xspace.}
    \end{subfigure}
    \hfill
    \begin{subfigure}[t]{0.3\textwidth}
        \centering
        \includegraphics[width=\textwidth]{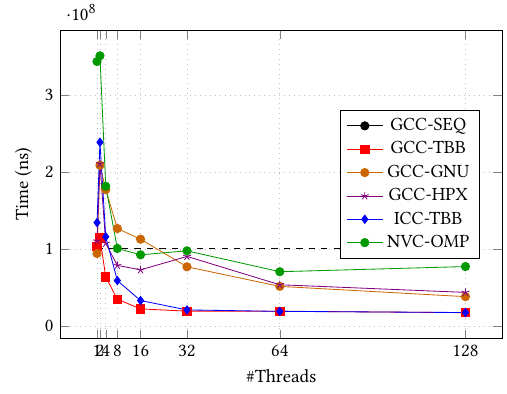}
        \caption{Execution time scalability with the number of threads used. Problem size is $2^{30}$\xspace. Lower is better\xspace.}
    \end{subfigure}
    \hfill
    \begin{subfigure}[t]{0.3\textwidth}
        \centering
        \includegraphics[width=\textwidth]{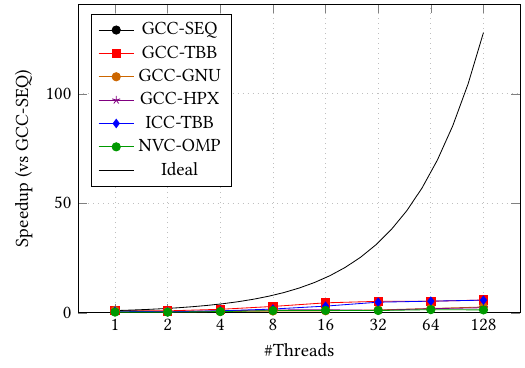}
        \caption{Speedup against GCC's seq. implementation. Problem size is $2^{30}$\xspace. Higher is better\xspace.}
    \end{subfigure}
    \caption{Results for benchmark \texttt{X::find}\xspace; \mbox{\emph{VSC-5}}\xspace.}
    \label{fig:vsc-find}
\end{figure*}

\FloatBarrier
\clearpage
\section{\texttt{X::for\_each}\xspace} \label{ap:foreach}
\begin{figure*}[h!]
    \centering
    \begin{subfigure}[t]{0.3\textwidth}
        \centering
        \includegraphics[width=\textwidth]{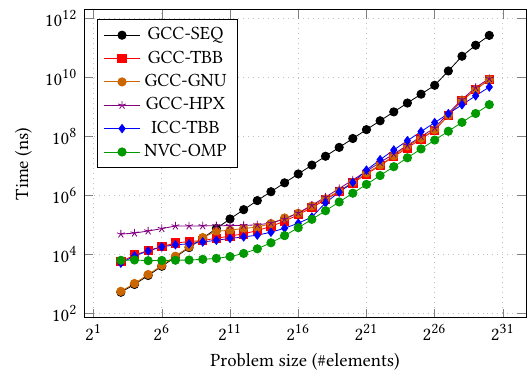}
        \caption{Execution time scalability with the problem size. All cores used except for GCC’s seq. implementation\xspace. Lower is better\xspace.}
    \end{subfigure}
    \hfill
    \begin{subfigure}[t]{0.3\textwidth}
        \centering
        \includegraphics[width=\textwidth]{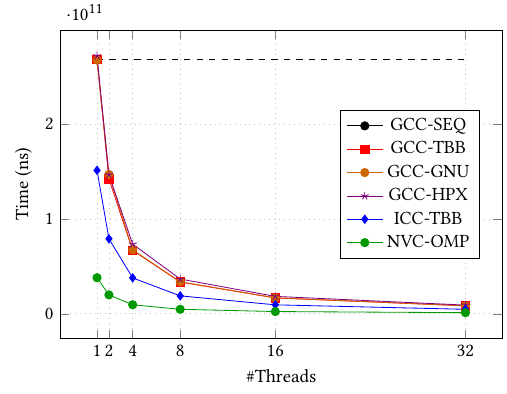}
        \caption{Execution time scalability with the number of threads used. Problem size is $2^{30}$\xspace. Lower is better\xspace.}
    \end{subfigure}
    \hfill
    \begin{subfigure}[t]{0.3\textwidth}
        \centering
        \includegraphics[width=\textwidth]{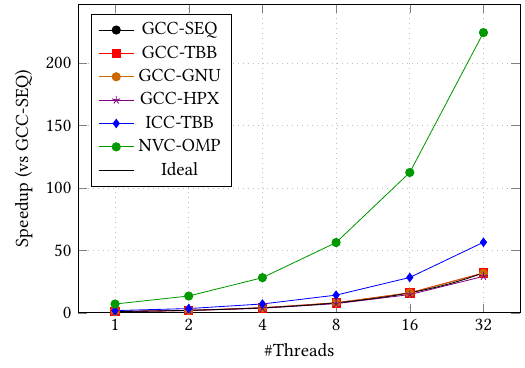}
        \caption{Speedup against GCC's seq. implementation. Problem size is $2^{30}$\xspace. Higher is better\xspace.}
    \end{subfigure}
    \caption{Results for benchmark \texttt{X::for\_each}\xspace; \emph{Hydra}\xspace.}
    \label{fig:hydra-foreach}
\end{figure*}

\begin{figure*}[h!]
    \centering
    \begin{subfigure}[t]{0.3\textwidth}
        \centering
        \includegraphics[width=\textwidth]{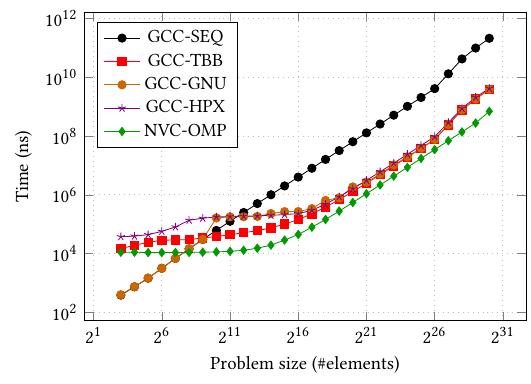}
        \caption{Execution time scalability with the problem size. All cores used except for GCC’s seq. implementation\xspace. Lower is better\xspace.}
    \end{subfigure}
    \hfill
    \begin{subfigure}[t]{0.3\textwidth}
        \centering
        \includegraphics[width=\textwidth]{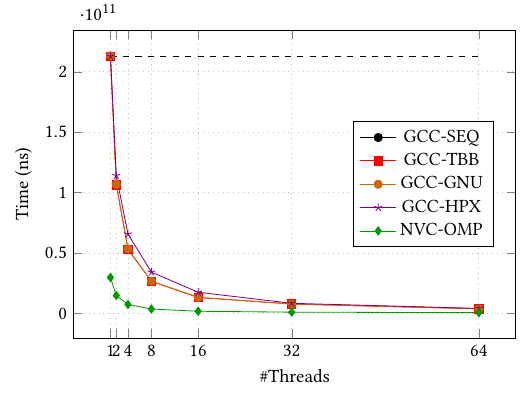}
        \caption{Execution time scalability with the number of threads used. Problem size is $2^{30}$\xspace. Lower is better\xspace.}
    \end{subfigure}
    \hfill
    \begin{subfigure}[t]{0.3\textwidth}
        \centering
        \includegraphics[width=\textwidth]{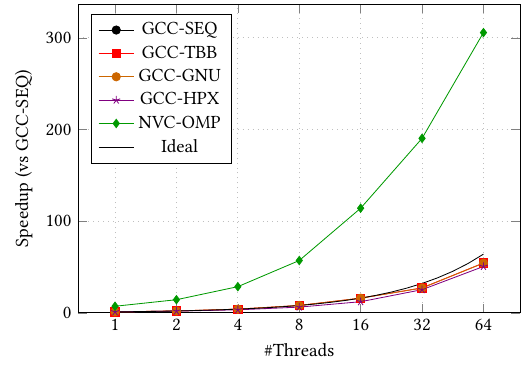}
        \caption{Speedup against GCC's seq. implementation. Problem size is $2^{30}$\xspace. Higher is better\xspace.}
    \end{subfigure}
    \caption{Results for benchmark \texttt{X::for\_each}\xspace; \emph{Nebula}\xspace.}
    \label{fig:nebula-foreach}
\end{figure*}

\begin{figure*}[h!]
    \centering
    \begin{subfigure}[t]{0.3\textwidth}
        \centering
        \includegraphics[width=\textwidth]{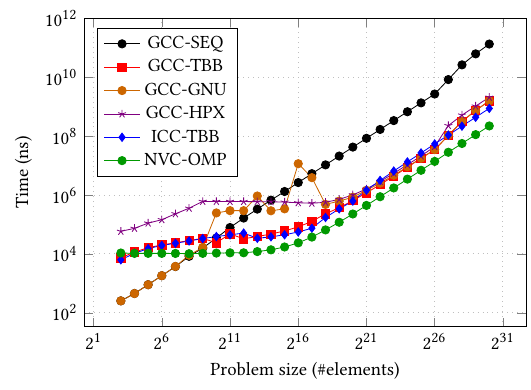}
        \caption{Execution time scalability with the problem size. All cores used except for GCC’s seq. implementation\xspace. Lower is better\xspace.}
    \end{subfigure}
    \hfill
    \begin{subfigure}[t]{0.3\textwidth}
        \centering
        \includegraphics[width=\textwidth]{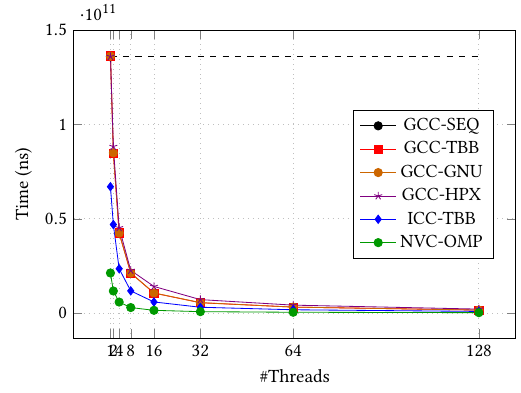}
        \caption{Execution time scalability with the number of threads used. Problem size is $2^{30}$\xspace. Lower is better\xspace.}
    \end{subfigure}
    \hfill
    \begin{subfigure}[t]{0.3\textwidth}
        \centering
        \includegraphics[width=\textwidth]{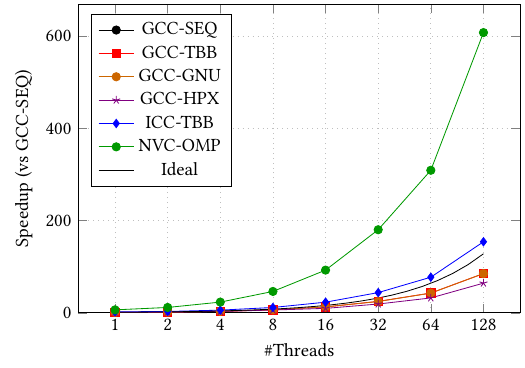}
        \caption{Speedup against GCC's seq. implementation. Problem size is $2^{30}$\xspace. Higher is better\xspace.}
    \end{subfigure}
    \caption{Results for benchmark \texttt{X::for\_each}\xspace; \mbox{\emph{VSC-5}}\xspace.}
    \label{fig:vsc-foreach}
\end{figure*}

\FloatBarrier
\clearpage
\section{\texttt{X::inclusive\_scan}\xspace} \label{ap:inclscan}
\begin{figure*}[h!]
    \centering
    \begin{subfigure}[t]{0.3\textwidth}
        \centering
        \includegraphics[width=\textwidth]{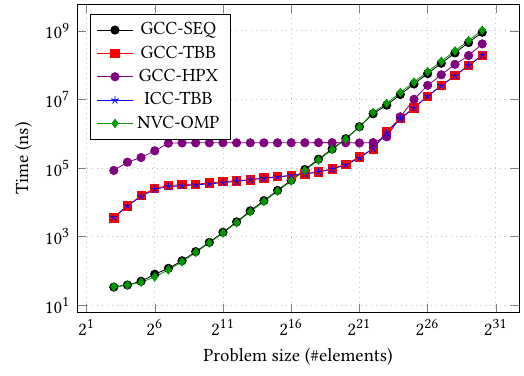}
        \caption{Execution time scalability with the problem size. All cores used except for GCC’s seq. implementation\xspace. Lower is better\xspace.}
    \end{subfigure}
    \hfill
    \begin{subfigure}[t]{0.3\textwidth}
        \centering
        \includegraphics[width=\textwidth]{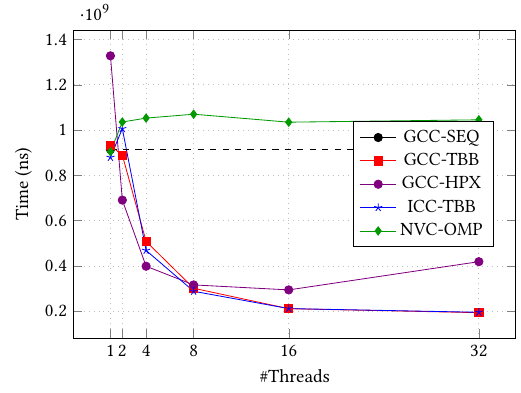}
        \caption{Execution time scalability with the number of threads used. Problem size is $2^{30}$\xspace. Lower is better\xspace.}
    \end{subfigure}
    \hfill
    \begin{subfigure}[t]{0.3\textwidth}
        \centering
        \includegraphics[width=\textwidth]{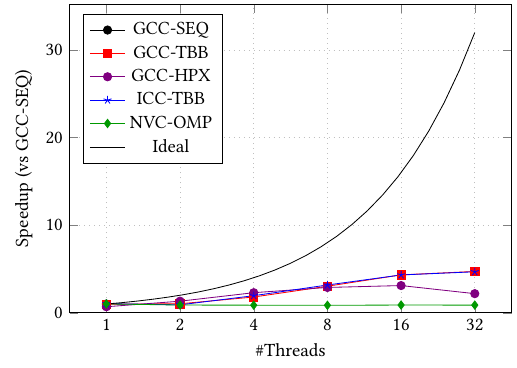}
        \caption{Speedup against GCC's seq. implementation. Problem size is $2^{30}$\xspace. Higher is better\xspace.}
    \end{subfigure}
    \caption{Results for benchmark \texttt{X::inclusive\_scan}\xspace; \emph{Hydra}\xspace.}
    \label{fig:hydra-incl-scan}
\end{figure*}

\begin{figure*}[h!]
    \centering
    \begin{subfigure}[t]{0.3\textwidth}
        \centering
        \includegraphics[width=\textwidth]{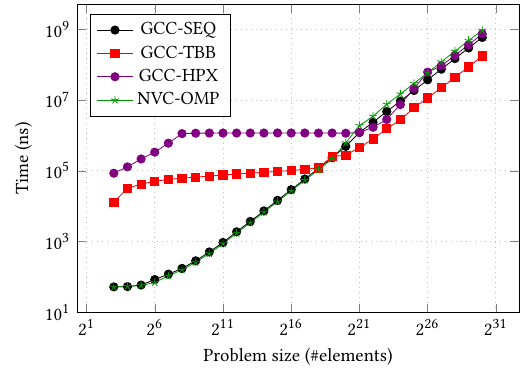}
        \caption{Execution time scalability with the problem size. All cores used except for GCC’s seq. implementation\xspace. Lower is better\xspace.}
    \end{subfigure}
    \hfill
    \begin{subfigure}[t]{0.3\textwidth}
        \centering
        \includegraphics[width=\textwidth]{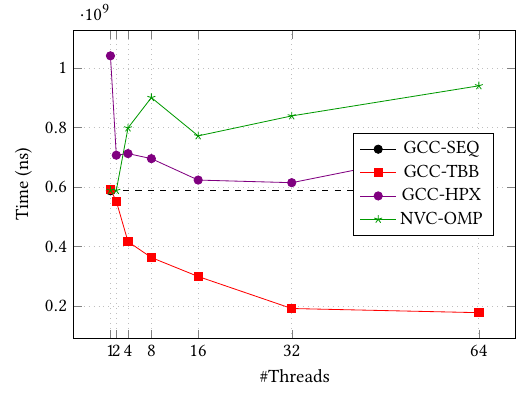}
        \caption{Execution time scalability with the number of threads used. Problem size is $2^{30}$\xspace. Lower is better\xspace.}
    \end{subfigure}
    \hfill
    \begin{subfigure}[t]{0.3\textwidth}
        \centering
        \includegraphics[width=\textwidth]{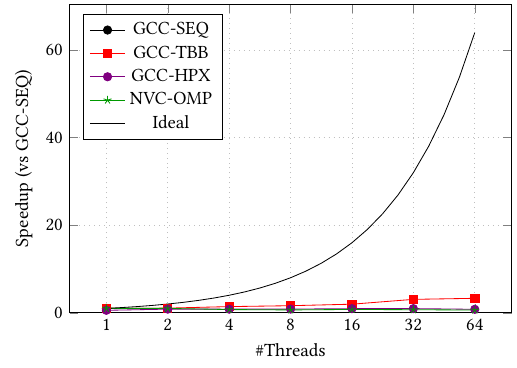}
        \caption{Speedup against GCC's seq. implementation. Problem size is $2^{30}$\xspace. Higher is better\xspace.}
    \end{subfigure}
    \caption{Results for benchmark \texttt{X::inclusive\_scan}\xspace; \emph{Nebula}\xspace.}
    \label{fig:nebula-incl-scan}
\end{figure*}

\begin{figure*}[h!]
    \centering
    \begin{subfigure}[t]{0.3\textwidth}
        \centering
        \includegraphics[width=\textwidth]{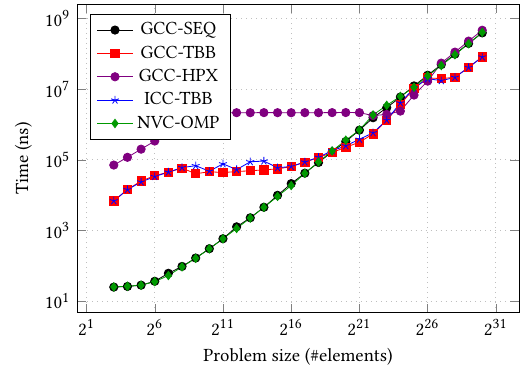}
        \caption{Execution time scalability with the problem size. All cores used except for GCC’s seq. implementation\xspace. Lower is better\xspace.}
    \end{subfigure}
    \hfill
    \begin{subfigure}[t]{0.3\textwidth}
        \centering
        \includegraphics[width=\textwidth]{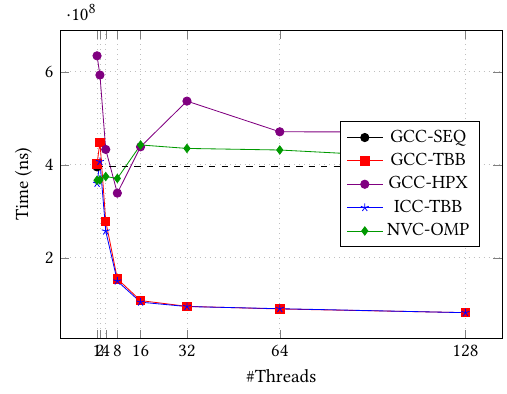}
        \caption{Execution time scalability with the number of threads used. Problem size is $2^{30}$\xspace. Lower is better\xspace.}
    \end{subfigure}
    \hfill
    \begin{subfigure}[t]{0.3\textwidth}
        \centering
        \includegraphics[width=\textwidth]{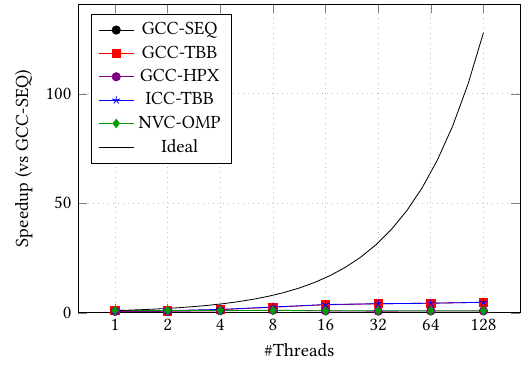}
        \caption{Speedup against GCC's seq. implementation. Problem size is $2^{30}$\xspace. Higher is better\xspace.}
    \end{subfigure}
    \caption{Results for benchmark \texttt{X::inclusive\_scan}\xspace; \mbox{\emph{VSC-5}}\xspace.}
    \label{fig:vsc-incl-scan}
\end{figure*}

\FloatBarrier
\clearpage
\section{\texttt{X::reduce}\xspace} \label{ap:reduce}
\begin{figure*}[h!]
    \centering
    \begin{subfigure}[t]{0.3\textwidth}
        \centering
        \includegraphics[width=\textwidth]{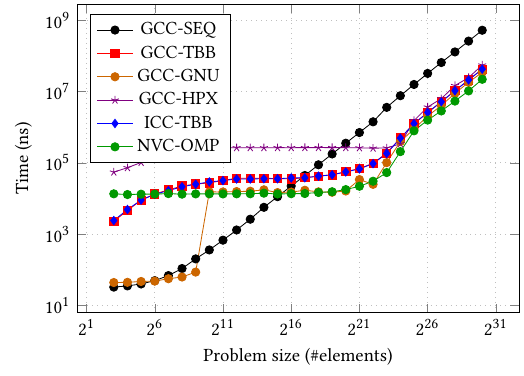}
        \caption{Execution time scalability with the problem size. All cores used except for GCC’s seq. implementation\xspace. Lower is better\xspace.}
    \end{subfigure}
    \hfill
    \begin{subfigure}[t]{0.3\textwidth}
        \centering
        \includegraphics[width=\textwidth]{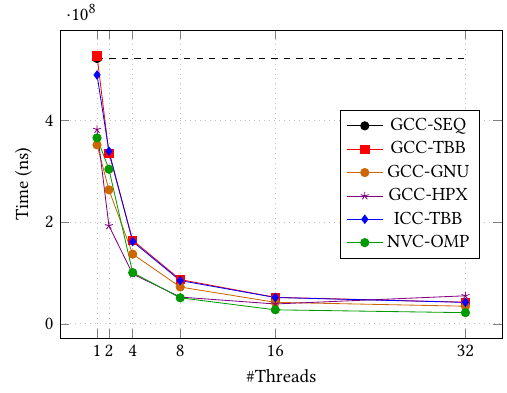}
        \caption{Execution time scalability with the number of threads used. Problem size is $2^{30}$\xspace. Lower is better\xspace.}
    \end{subfigure}
    \hfill
    \begin{subfigure}[t]{0.3\textwidth}
        \centering
        \includegraphics[width=\textwidth]{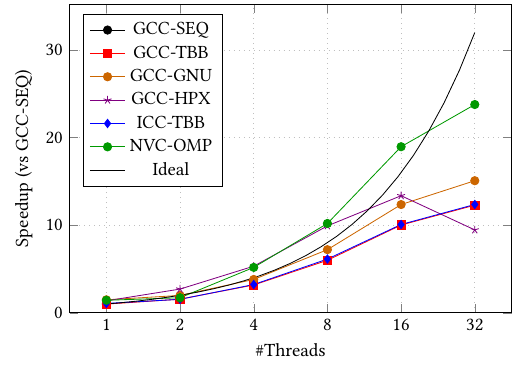}
        \caption{Speedup against GCC's seq. implementation. Problem size is $2^{30}$\xspace. Higher is better\xspace.}
    \end{subfigure}
    \caption{Results for benchmark \texttt{X::reduce}\xspace; \emph{Hydra}\xspace.}
    \label{fig:hydra-reduce}
\end{figure*}

\begin{figure*}[h!]
    \centering
    \begin{subfigure}[t]{0.3\textwidth}
        \centering
        \includegraphics[width=\textwidth]{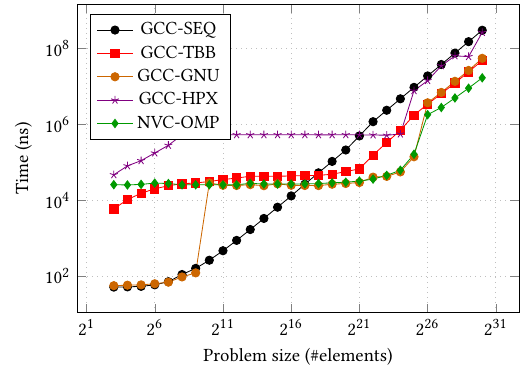}
        \caption{Execution time scalability with the problem size. All cores used except for GCC’s seq. implementation\xspace. Lower is better\xspace.}
    \end{subfigure}
    \hfill
    \begin{subfigure}[t]{0.3\textwidth}
        \centering
        \includegraphics[width=\textwidth]{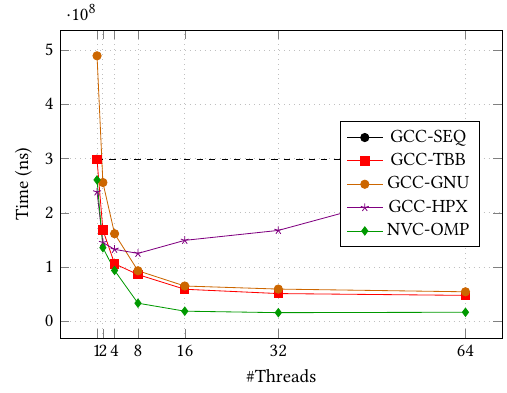}
        \caption{Execution time scalability with the number of threads used. Problem size is $2^{30}$\xspace. Lower is better\xspace.}
    \end{subfigure}
    \hfill
    \begin{subfigure}[t]{0.3\textwidth}
        \centering
        \includegraphics[width=\textwidth]{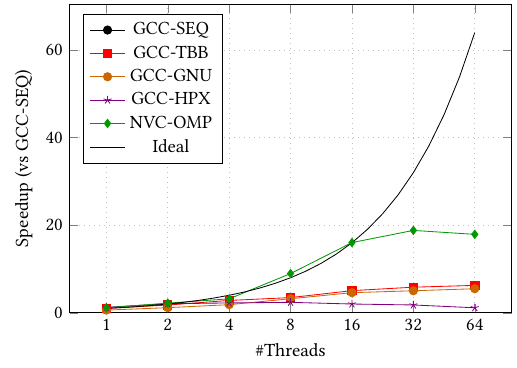}
        \caption{Speedup against GCC's seq. implementation. Problem size is $2^{30}$\xspace. Higher is better\xspace.}
    \end{subfigure}
    \caption{Results for benchmark \texttt{X::reduce}\xspace; \emph{Nebula}\xspace.}
    \label{fig:nebula-reduce}
\end{figure*}

\begin{figure*}[h!]
    \centering
    \begin{subfigure}[t]{0.3\textwidth}
        \centering
        \includegraphics[width=\textwidth]{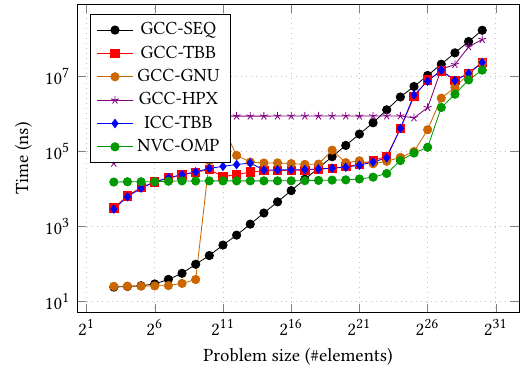}
        \caption{Execution time scalability with the problem size. All cores used except for GCC’s seq. implementation\xspace. Lower is better\xspace.}
    \end{subfigure}
    \hfill
    \begin{subfigure}[t]{0.3\textwidth}
        \centering
        \includegraphics[width=\textwidth]{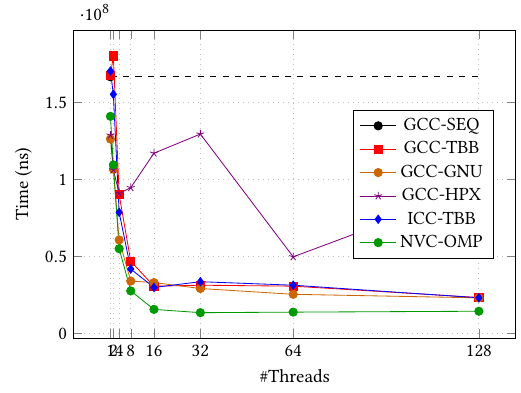}
        \caption{Execution time scalability with the number of threads used. Problem size is $2^{30}$\xspace. Lower is better\xspace.}
    \end{subfigure}
    \hfill
    \begin{subfigure}[t]{0.3\textwidth}
        \centering
        \includegraphics[width=\textwidth]{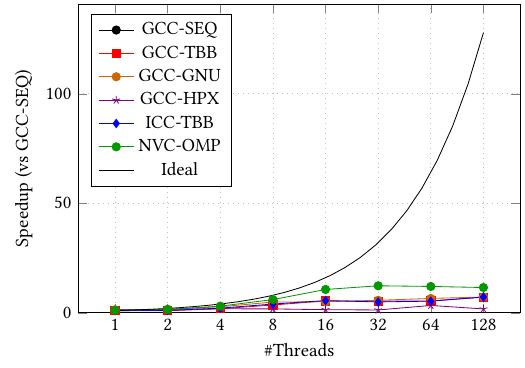}
        \caption{Speedup against GCC's seq. implementation. Problem size is $2^{30}$\xspace. Higher is better\xspace.}
    \end{subfigure}
    \caption{Results for benchmark \texttt{X::reduce}\xspace; \mbox{\emph{VSC-5}}\xspace.}
    \label{fig:vsc-reduce}
\end{figure*}

\FloatBarrier
\clearpage
\section{\texttt{X::sort}\xspace} \label{ap:sort}
\begin{figure*}[h!]
    \centering
    \begin{subfigure}[t]{0.3\textwidth}
        \centering
        \includegraphics[width=\textwidth]{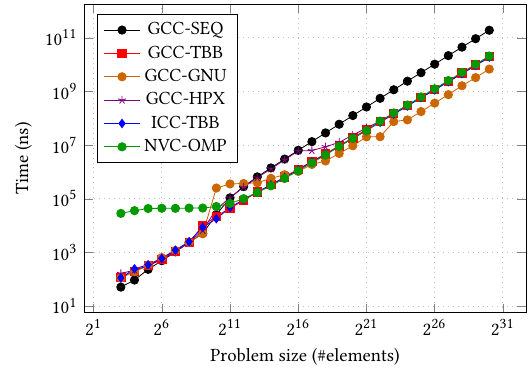}
        \caption{Execution time scalability with the problem size. All cores used except for GCC’s seq. implementation\xspace. Lower is better\xspace.}
    \end{subfigure}
    \hfill
    \begin{subfigure}[t]{0.3\textwidth}
        \centering
        \includegraphics[width=\textwidth]{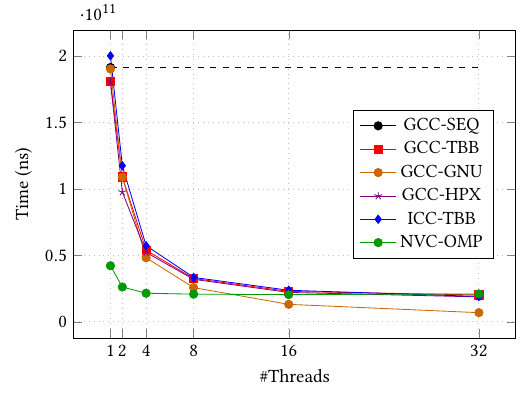}
        \caption{Execution time scalability with the number of threads used. Problem size is $2^{30}$\xspace. Lower is better\xspace.}
    \end{subfigure}
    \hfill
    \begin{subfigure}[t]{0.3\textwidth}
        \centering
        \includegraphics[width=\textwidth]{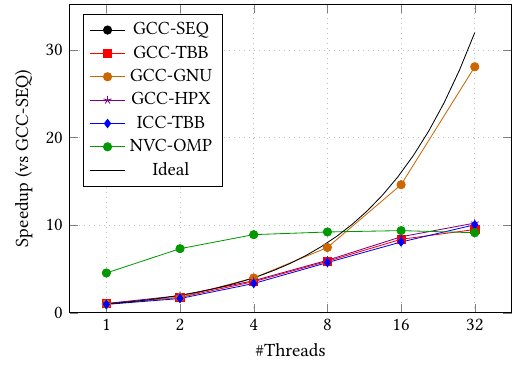}
        \caption{Speedup against GCC's seq. implementation. Problem size is $2^{30}$\xspace. Higher is better\xspace.}
    \end{subfigure}
    \caption{Results for benchmark \texttt{X::sort}\xspace; \emph{Hydra}\xspace.}
    \label{fig:hydra-sort}
\end{figure*}

\begin{figure*}[h!]
    \centering
    \begin{subfigure}[t]{0.3\textwidth}
        \centering
        \includegraphics[width=\textwidth]{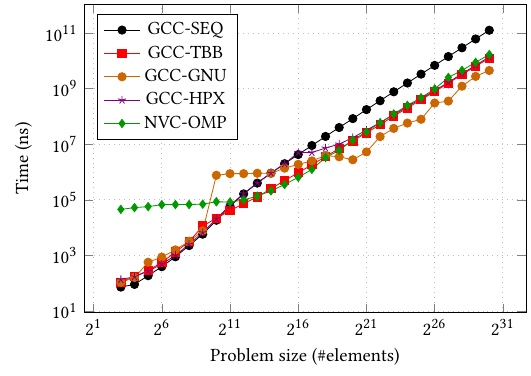}
        \caption{Execution time scalability with the problem size. All cores used except for GCC’s seq. implementation\xspace. Lower is better\xspace.}
    \end{subfigure}
    \hfill
    \begin{subfigure}[t]{0.3\textwidth}
        \centering
        \includegraphics[width=\textwidth]{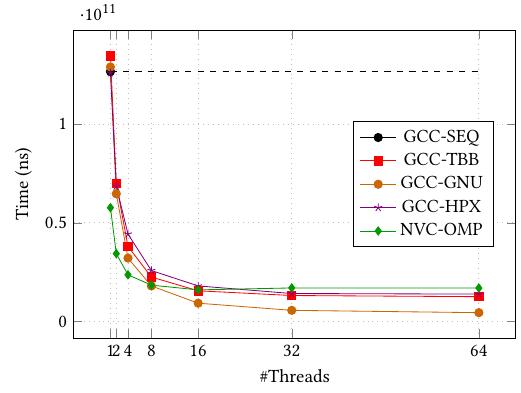}
        \caption{Execution time scalability with the number of threads used. Problem size is $2^{30}$\xspace. Lower is better\xspace.}
    \end{subfigure}
    \hfill
    \begin{subfigure}[t]{0.3\textwidth}
        \centering
        \includegraphics[width=\textwidth]{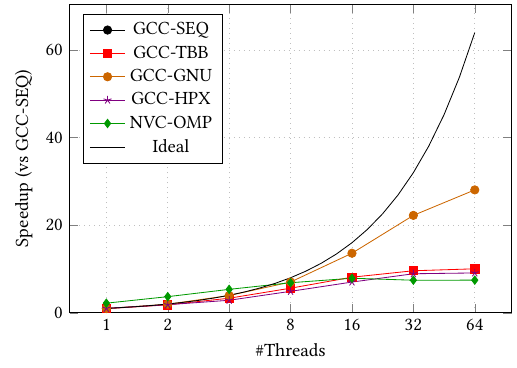}
        \caption{Speedup against GCC's seq. implementation. Problem size is $2^{30}$\xspace. Higher is better\xspace.}
    \end{subfigure}
    \caption{Results for benchmark \texttt{X::sort}\xspace; \emph{Nebula}\xspace.}
    \label{fig:nebula-sort}
\end{figure*}

\begin{figure*}[h!]
    \centering
    \begin{subfigure}[t]{0.3\textwidth}
        \centering
        \includegraphics[width=\textwidth]{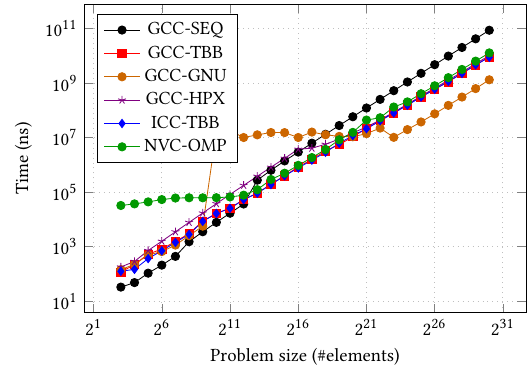}
        \caption{Execution time scalability with the problem size. All cores used except for GCC’s seq. implementation\xspace. Lower is better\xspace.}
    \end{subfigure}
    \hfill
    \begin{subfigure}[t]{0.3\textwidth}
        \centering
        \includegraphics[width=\textwidth]{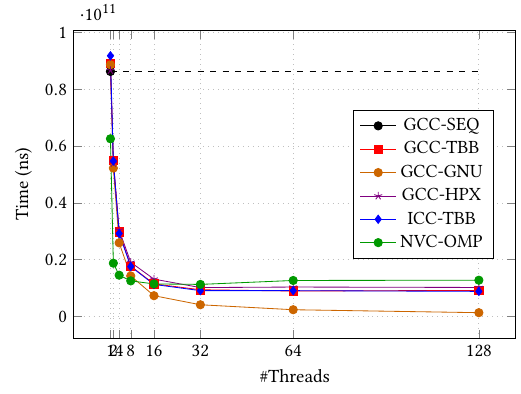}
        \caption{Execution time scalability with the number of threads used. Problem size is $2^{30}$\xspace. Lower is better\xspace.}
    \end{subfigure}
    \hfill
    \begin{subfigure}[t]{0.3\textwidth}
        \centering
        \includegraphics[width=\textwidth]{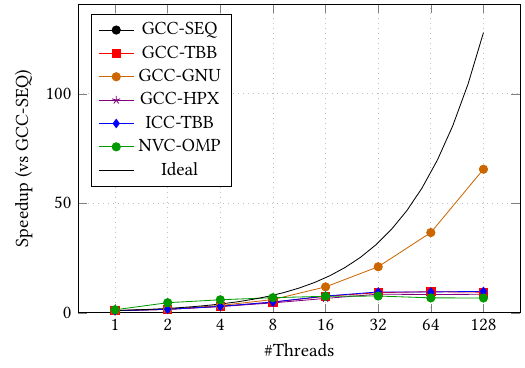}
        \caption{Speedup against GCC's seq. implementation. Problem size is $2^{30}$\xspace. Higher is better\xspace.}
    \end{subfigure}
    \caption{Results for benchmark \texttt{X::sort}\xspace; \mbox{\emph{VSC-5}}\xspace.}
    \label{fig:vsc-sort}
\end{figure*}

\end{document}